\documentclass{article}
\usepackage{array}
\usepackage{listings}
\usepackage{upquote}
\usepackage[noadjust]{cite}
\usepackage{graphicx}
\usepackage{float}
\usepackage[hidelinks]{hyperref}
\usepackage[font=footnotesize, labelfont=footnotesize]{caption}
\usepackage{tikz}
\usetikzlibrary{positioning}
\title{SGL: A Structured Graphics Language}
\author{Jon Chapman \\ \footnotesize{jochapjo@proton.me}}
\date{}
\begin{document}
\maketitle
\begin{abstract}
This paper introduces SGL, a graphics language that is aesthetically similar to SQL.
As a graphical counterpart to SQL, SGL enables specification of statistical graphics within SQL query interfaces. 
SGL is based on a grammar of graphics that has been customized to support a SQL aesthetic.

This paper presents the fundamental components of the SGL language alongside examples,
and describes SGL's underlying grammar of graphics via comparison to its closest predecessor,
the layered grammar of graphics.
\end{abstract}

\section{Introduction}

High-level grammars of graphics enable concise yet expressive specification within the context of statistical graphics, making them well-suited to exploratory data analysis.
Since Wilkinson first formalized such a grammar in \emph{The Grammar of Graphics} \cite{wilkinson:2005}, several implementations (often with modification to Wilkinson's original grammar) have gained popularity.
These implementations have embedded grammars in various environments, allowing users to specify graphics using GUI's (e.g. Tableau, formerly Polaris \cite{tableau:2002}), programming languages (e.g. R via ggplot2 \cite{wickham:2010}), and language-independent data formats (e.g. JSON via Vega-Lite \cite{vegalite:2017}).

However, no implementation has incorporated such a grammar into an interactive SQL query interface.
These interfaces are a common environment for data exploration, making visualization an essential, yet often lacking, capability.
To remedy this, it is beneficial to develop a graphics language suitable for this environment.

This paper introduces SGL, a graphics language that is aesthetically similar to SQL.
SGL is based on a grammar of graphics that has been adapted from prior grammars \cite{wickham:2010, wilkinson:2005, vegalite:2017} to support the SQL-like characteristics of the language.
The similarity to SQL and the concise yet expressive nature derived from the grammatical foundation make it a suitable graphical counterpart to SQL.

This paper presents the fundamental components of the language alongside example SGL statements.
Additionally, SGL's underlying grammar of graphics is elaborated via comparison to the layered grammar of graphics \cite{wickham:2010}.

\section{The SGL Language}
To introduce the language, we assume that SGL is operating within a data warehouse, and that this warehouse contains tables named \texttt{cars} and \texttt{trees}.
Each row in these tables represents an instance of an object (a car or tree), and the columns represent attributes of the object.
Figures \ref{fig:cars-sample} and \ref{fig:trees-sample} show samples of data from these tables.

\begin{figure}[H]
\centering
\begin{minipage}[t]{0.15\textwidth}
\vspace*{10pt}
\lstset{language=SQL, basicstyle=\ttfamily, columns=flexible}
\begin{lstlisting}
select *
from cars
limit 5;
\end{lstlisting}
\end{minipage}
\hfill
\begin{minipage}[t]{0.75\textwidth}
\vspace*{0pt}
\raggedright
\small
\begin{tabular}{|>{\ttfamily}l|>{\ttfamily}l|>{\ttfamily}l|>{\ttfamily}l|>{\ttfamily}l|}
\hline
\textbf{car\_id} & \textbf{horsepower} & \textbf{miles\_per\_gallon} & \textbf{origin} & \textbf{year} \\
\hline
1 & 130 & 18 & USA & 1970 \\
2 & 165 & 15 & USA & 1970 \\
3 & 150 & 18 & USA & 1970 \\
4 & 150 & 16 & USA & 1970 \\
5 & 140 & 17 & USA & 1970 \\
\hline
\end{tabular}
\end{minipage}
\caption{SQL statement and corresponding result set showing a sample of data from the \texttt{cars} table.}
\label{fig:cars-sample}
\end{figure}

\begin{figure}[H]
\centering
\begin{minipage}[t]{0.25\textwidth}
\vspace*{10pt}
\lstset{language=SQL, basicstyle=\ttfamily, columns=flexible}
\begin{lstlisting}
select *
from trees
limit 5;
\end{lstlisting}
\end{minipage}
\hfill
\begin{minipage}[t]{0.65\textwidth}
\vspace*{0pt}
\raggedright
\small
\begin{tabular}{|>{\ttfamily}l|>{\ttfamily}l|>{\ttfamily}l|}
\hline
\textbf{tree\_id} & \textbf{age} & \textbf{circumference} \\
\hline
1 & 118 & 30 \\
1 & 484 & 58 \\
1 & 664 & 87 \\
1 & 1004 & 115 \\
1 & 1231 & 120 \\
\hline
\end{tabular}
\end{minipage}
\caption{SQL statement and corresponding result set showing a sample of data from the \texttt{trees} table.}
\label{fig:trees-sample}
\end{figure}

\subsection{The From Clause}
\begin{figure}[H]
\centering
\begin{minipage}[t]{0.35\textwidth}
\vspace*{42pt}
\lstset{language=SQL, basicstyle=\ttfamily, columns=flexible}
\begin{lstlisting}
visualize
  horsepower as x,
  miles_per_gallon as y
from cars
using points;
\end{lstlisting}
\end{minipage}
\hfill
\begin{minipage}[t]{0.55\textwidth}
\vspace*{0pt}
\raggedleft
\includegraphics[width=0.9\textwidth]{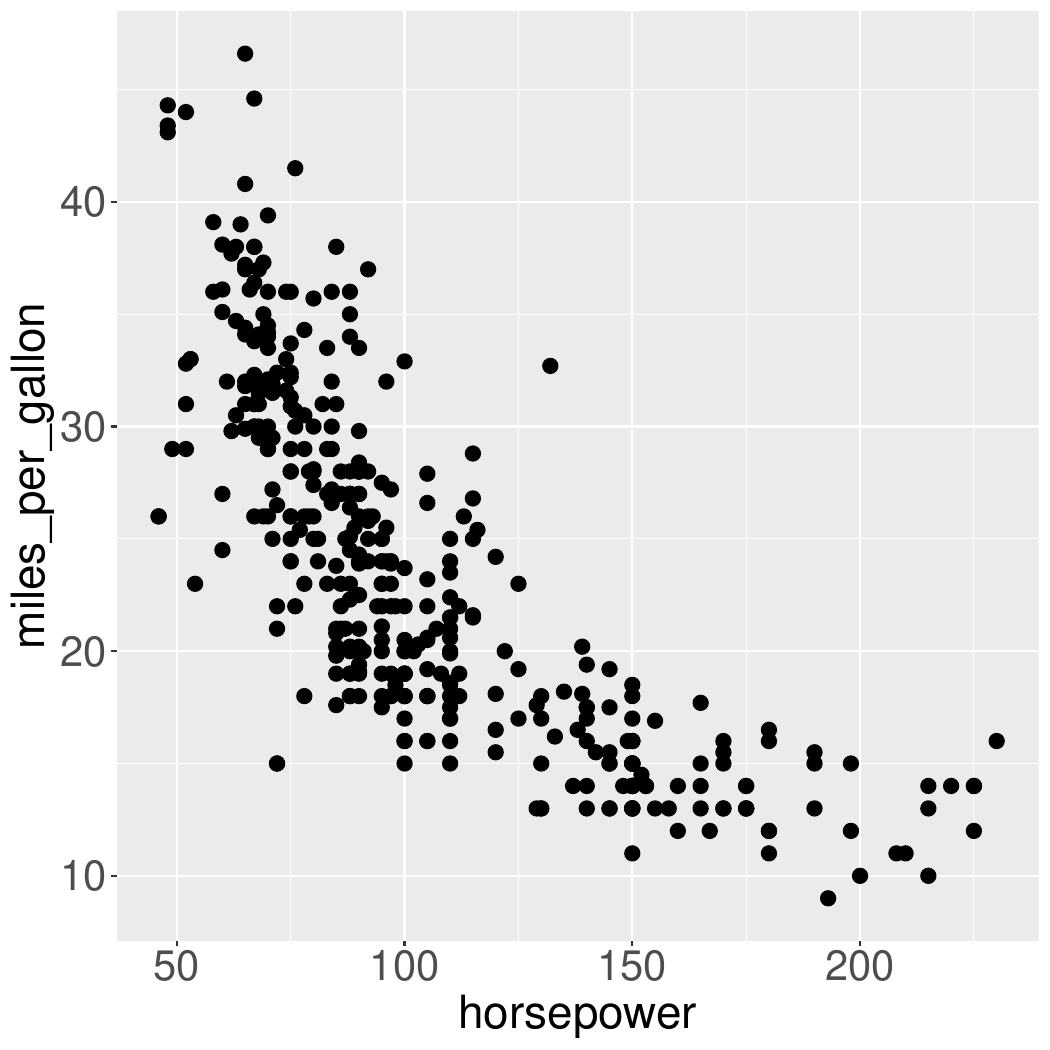}
\end{minipage}
\caption{SGL statement with a \texttt{from} clause that designates the \texttt{cars} table as a data source.}
\label{fig:from-table-source}
\end{figure}

The \texttt{from} keyword precedes a data source specification, which is often the name of a table or view in the data warehouse. Figure \ref{fig:from-table-source} shows a SGL statement
and corresponding graphic that specifies the \texttt{cars} table as the data source.

\begin{figure}[H]
\centering
\begin{minipage}[t]{0.35\textwidth}
\vspace*{18pt}
\lstset{language=SQL, basicstyle=\ttfamily, columns=flexible}
\begin{lstlisting}
visualize
  horsepower as x,
  miles_per_gallon as y
from (
  select *
  from cars
  where origin = 'Japan'
)
using points;
\end{lstlisting}
\end{minipage}
\hfill
\begin{minipage}[t]{0.55\textwidth}
\vspace*{0pt}
\includegraphics[width=0.9\textwidth]{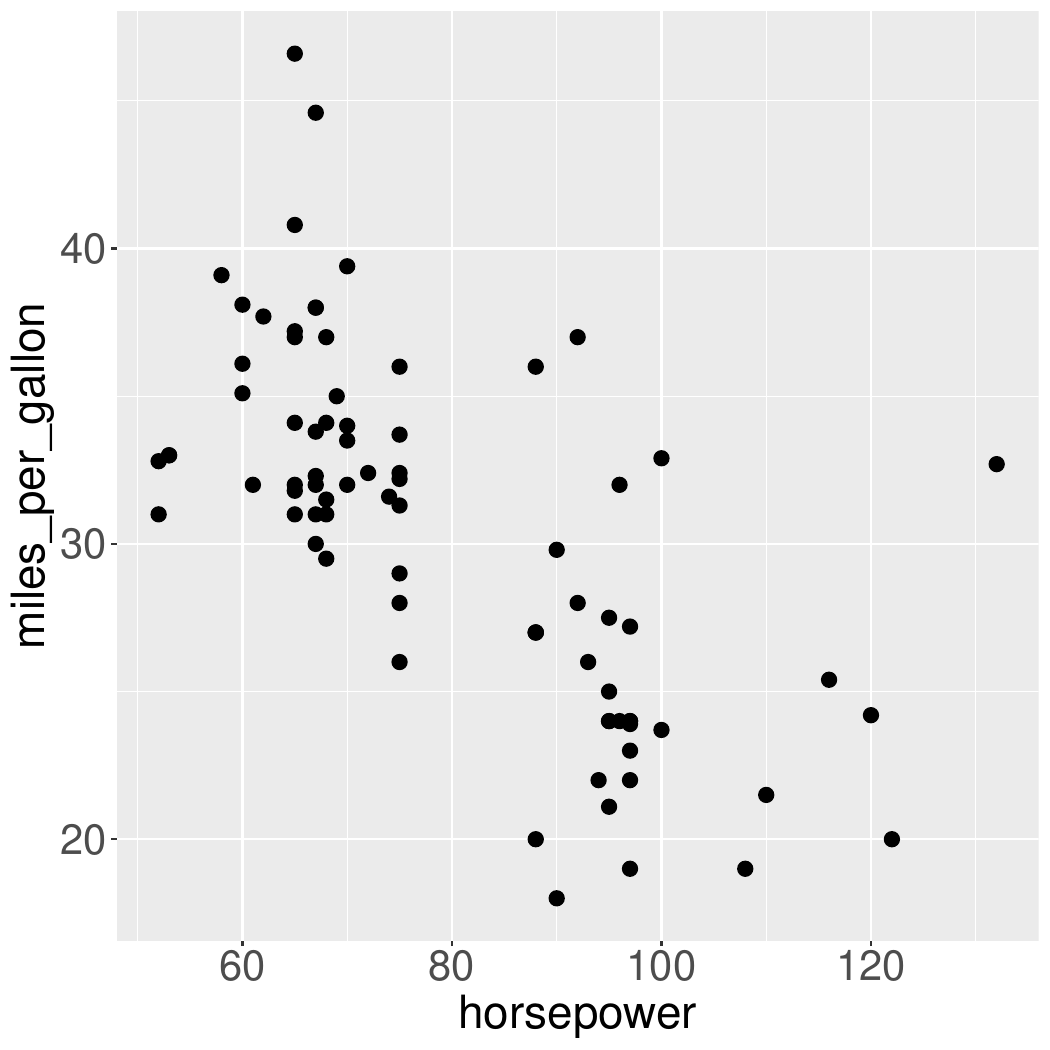}
\end{minipage}
\caption{SGL statement with a SQL subquery as the data source, resulting in a graphic similar to Figure \ref{fig:from-table-source}, except that only Japanese cars are included.}
\label{fig:from-subquery-source}
\end{figure}

This is similar in usage to the \texttt{from} keyword in SQL, except that only a single data source is allowed (i.e., a comma-separated list of table names is not valid).
 If data from multiple sources or pre-processing of data is necessary, then a SQL subquery can be provided, as shown in Figure \ref{fig:from-subquery-source}.

\subsection{The Using Clause}
The \texttt{using} keyword precedes the name of the geometric object(s) that will represent the data.
Following ggplot2 terminology \cite{wickham:2010}, these geometric objects are referred to as geoms.
Geom names (both the plural and singular forms) are keywords. Figures \ref{fig:from-table-source} and \ref{fig:from-subquery-source} demonstrate using the \texttt{points} keyword to specify that data should be represented by point geoms.

\subsection{The Visualize Clause}
\begin{figure}[H]
\centering
\begin{minipage}[t]{0.35\textwidth}
\vspace*{40pt}
\lstset{language=SQL, basicstyle=\ttfamily, columns=flexible}
\begin{lstlisting}
visualize
  horsepower as x,
  miles_per_gallon as y,
  origin as color
from cars
using points;
\end{lstlisting}
\end{minipage}
\hfill
\begin{minipage}[t]{0.55\textwidth}
\vspace*{0pt}
\includegraphics[width=0.9\textwidth]{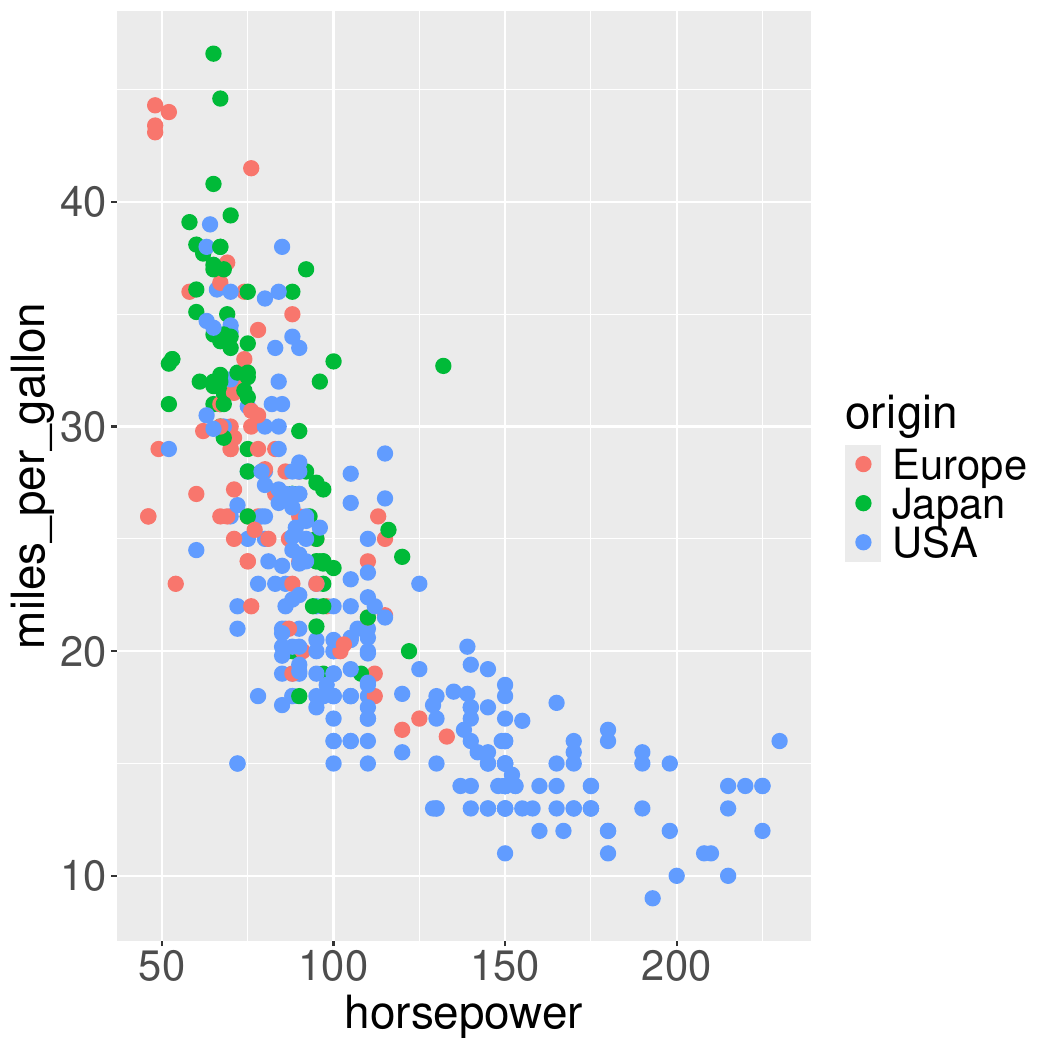}
\end{minipage}
\caption{SGL statement with a non-positional aesthetic \texttt{color}, resulting in point geoms that
are colored according to their \texttt{origin}.}
\label{fig:visualize-clause}
\end{figure}

The \texttt{visualize} keyword precedes the aesthetic-to-column mapping, which maps perceivable traits of the geom(s) to data source columns.
For example, Figure \ref{fig:from-table-source} maps the \texttt{x} and \texttt{y} positions of the point geoms to the \texttt{horsepower} and \texttt{miles\_per\_gallon} columns, respectively. Aesthetic names are keywords of the language,
and aesthetics may be non-positional, as shown in Figure \ref{fig:visualize-clause}.
The \texttt{visualize} keyword most closely resembles the \texttt{select} keyword within SQL.

In addition to mapping aesthetics to columns, aesthetics can be mapped to expressions that include transformations and aggregations.

\subsection{Column-Level Transformations and Aggregations}
\label{sec:column-level-transformations-and-aggregations}
\begin{figure}[H]
\centering
\begin{minipage}[t]{0.35\textwidth}
\vspace*{32pt}
\lstset{language=SQL, basicstyle=\ttfamily, columns=flexible}
\begin{lstlisting}
visualize
  bin(miles_per_gallon) as x,
  count(*) as y
from cars
group by
  bin(miles_per_gallon)
using bars;
\end{lstlisting}
\end{minipage}
\hfill
\begin{minipage}[t]{0.55\textwidth}
\vspace*{0pt}
\includegraphics[width=0.9\textwidth]{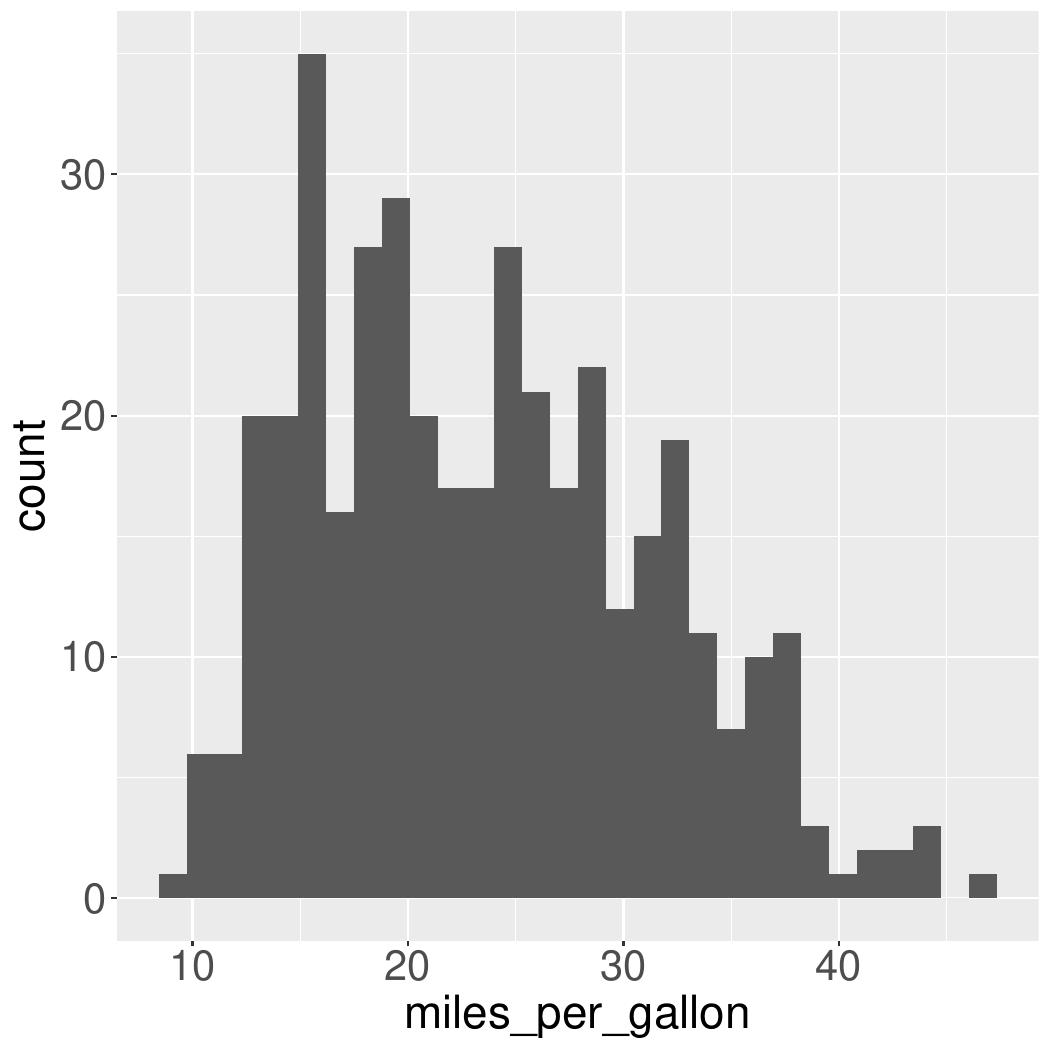}
\end{minipage}
\caption{SGL statement with a binning transformation and a count aggregation, resulting in a histogram on \texttt{miles\_per\_gallon}.}
\label{fig:bin-and-count}
\end{figure}

SGL supports column-level transformations and aggregations, as shown in Figure \ref{fig:bin-and-count} where a binning transformation is combined with a 
count aggregation to produce a histogram on \texttt{miles\_per\_gallon}. 

Here we see that, similarly to SQL, SGL has a \texttt{group by} clause where aggregation groupings are specified.
Any non-aggregated expressions in the \texttt{visualize} clause must also be included in the \texttt{group by} clause.
This is analogous to SQL's requirement that non-aggregated expressions in the \texttt{select} clause be included in SQL's \texttt{group by} clause.
Additional grouping on expressions not included in the \texttt{visualize} clause is allowed.

\begin{figure}[H]
\centering
\begin{minipage}[t]{0.35\textwidth}
\vspace*{20pt}
\lstset{language=SQL, basicstyle=\ttfamily, columns=flexible}
\begin{lstlisting}
visualize
  bin(miles_per_gallon) as x,
  count(*) as y
from cars
group by 
  bin(miles_per_gallon)
using bars
scale by
  log(x);
\end{lstlisting}
\end{minipage}
\hfill
\begin{minipage}[t]{0.55\textwidth}
\vspace*{0pt}
\includegraphics[width=0.9\textwidth]{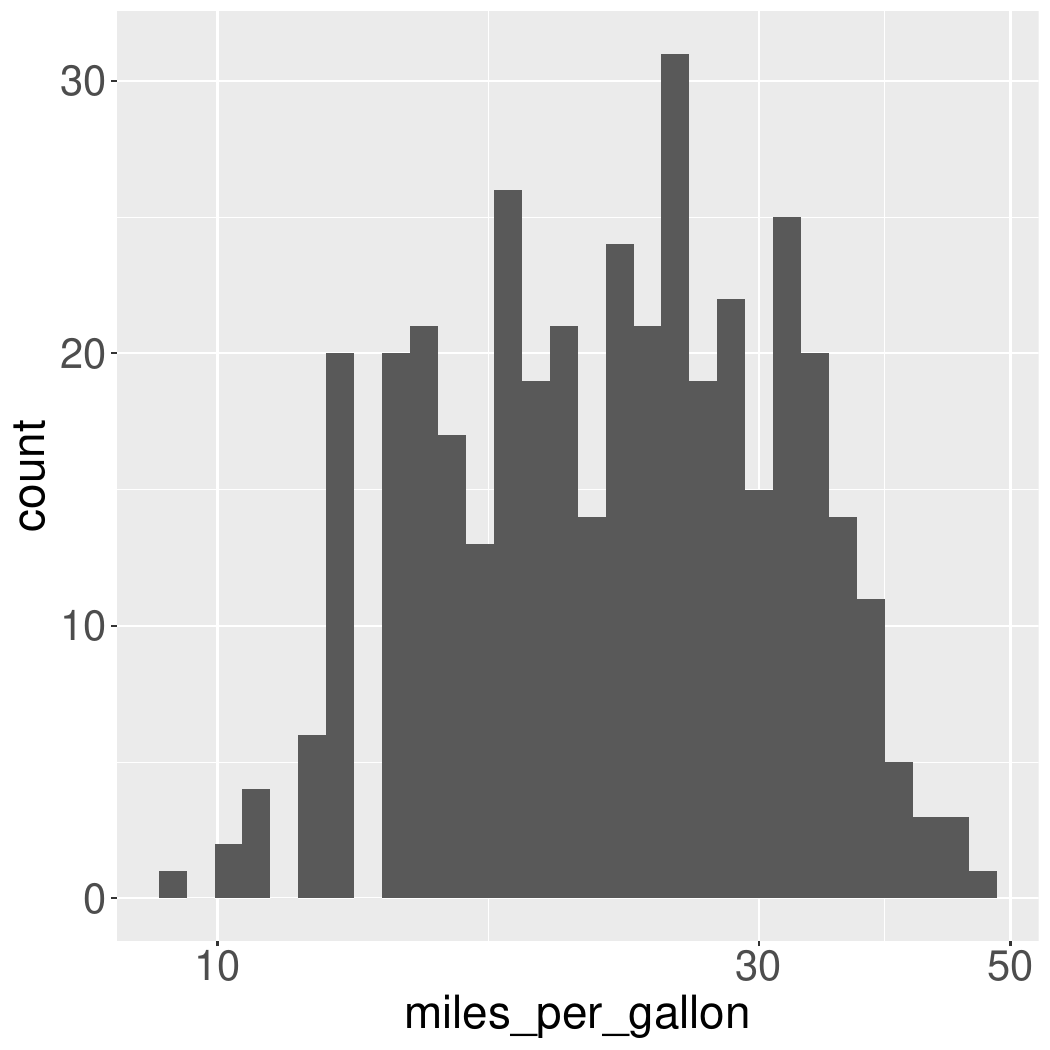}
\end{minipage}
\caption{SGL statement with a binning transformation that is performed after log-scaling, resulting in a log-scaled histogram on \texttt{miles\_per\_gallon}.}
\label{fig:log-scaled-histogram}
\end{figure}

Although SQL supports column-level transformation, grouping, and aggregation, it is desirable for SGL to provide additional support for these operations.
Statistical graphics often require operations such as binning that are not natively supported by SQL.
Additionally, SGL's column-level transformations and aggregations are performed after scaling, which cannot easily be replicated using SQL.
Figure \ref{fig:log-scaled-histogram} provides an example of this feature, where binning and counting are applied after log scaling, resulting in a log-scaled histogram. Scaling will be discussed further in section \ref{sec:the-scale-by-clause}.

Although we have presented column-level transformations and aggregations in conjunction, neither is a pre-requisite for the other, i.e., aggregations can be performed via grouping on columns that have no transformation,
and column-level transformations can be applied without any corresponding aggregation.

\subsection{The Collect By Clause}
\emph{ggplot2: Elegant Graphics for Data Analysis} \cite{wickham:2016} introduced a classification of geoms based on collectivity, where
geoms that represent each record by a distinct geometric object are classified as individual geoms, and geoms that represent multiple records by one
geometric object are classified as collective geoms.
To adapt this classification scheme to SGL's grammar of graphics, we first define the \textit{post-CTA dataset} to be the implicit dataset that results after column-level transformations and aggregations are applied.
If no column-level transformations or aggregations are specified in the SGL statement, then the post-CTA dataset is the same as the data source specified in the \texttt{from} clause.

\begin{figure}[H]
\centering
\begin{minipage}[t]{0.35\textwidth}
\vspace*{30pt}
\lstset{language=SQL, basicstyle=\ttfamily, columns=flexible}
\begin{lstlisting}
visualize
  year as x,
  mean(miles_per_gallon) as y
from cars
group by
  year
using points;
\end{lstlisting}
\end{minipage}
\hfill
\begin{minipage}[t]{0.55\textwidth}
\vspace*{0pt}
\includegraphics[width=0.9\textwidth]{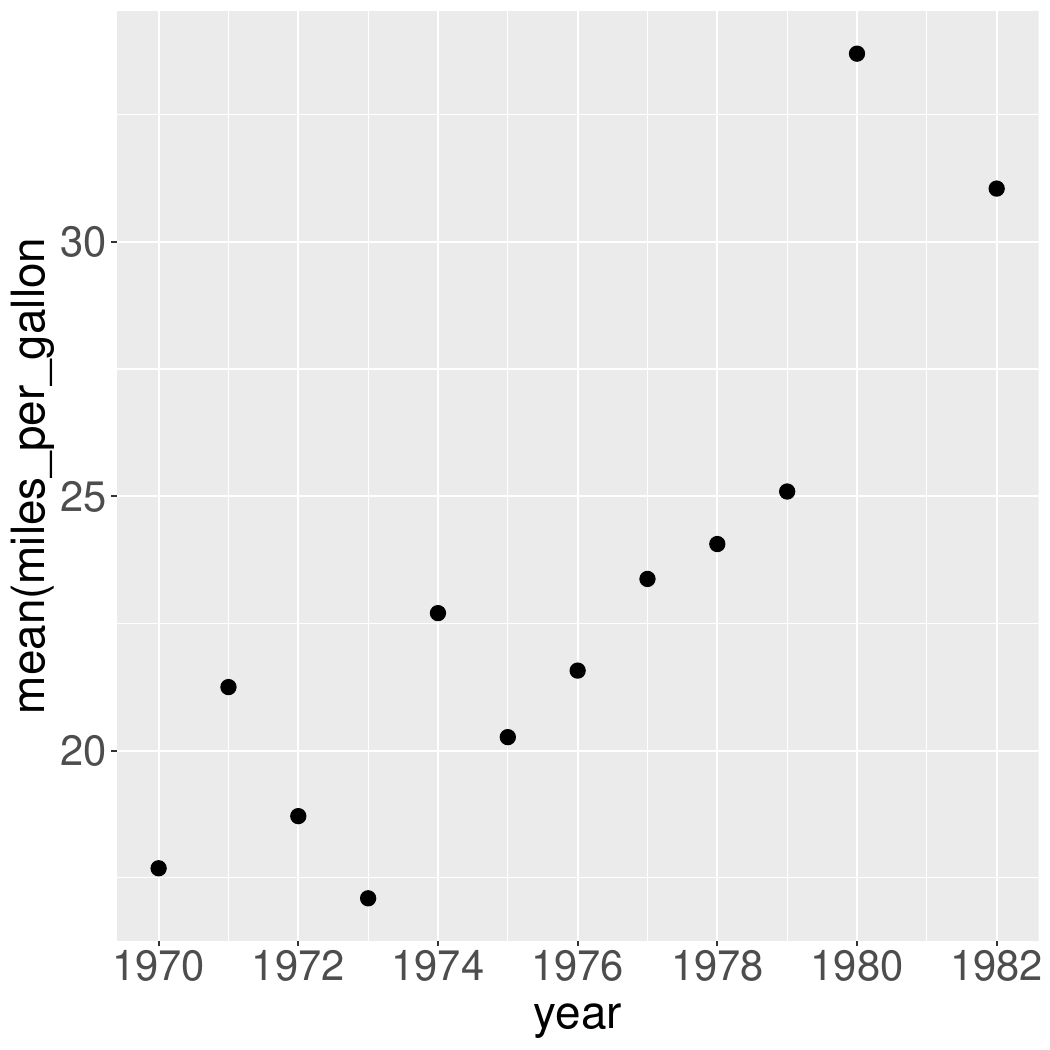}
\end{minipage}
\caption{SGL statement where each post-CTA record is represented by a distinct point object.}
\label{fig:individual-geom}
\end{figure}

\begin{figure}[H]
\centering
\begin{minipage}[t]{0.35\textwidth}
\vspace*{30pt}
\lstset{language=SQL, basicstyle=\ttfamily, columns=flexible}
\begin{lstlisting}
visualize
  year as x,
  mean(miles_per_gallon) as y
from cars
group by
  year
using line;
\end{lstlisting}
\end{minipage}
\hfill
\begin{minipage}[t]{0.55\textwidth}
\vspace*{0pt}
\includegraphics[width=0.9\textwidth]{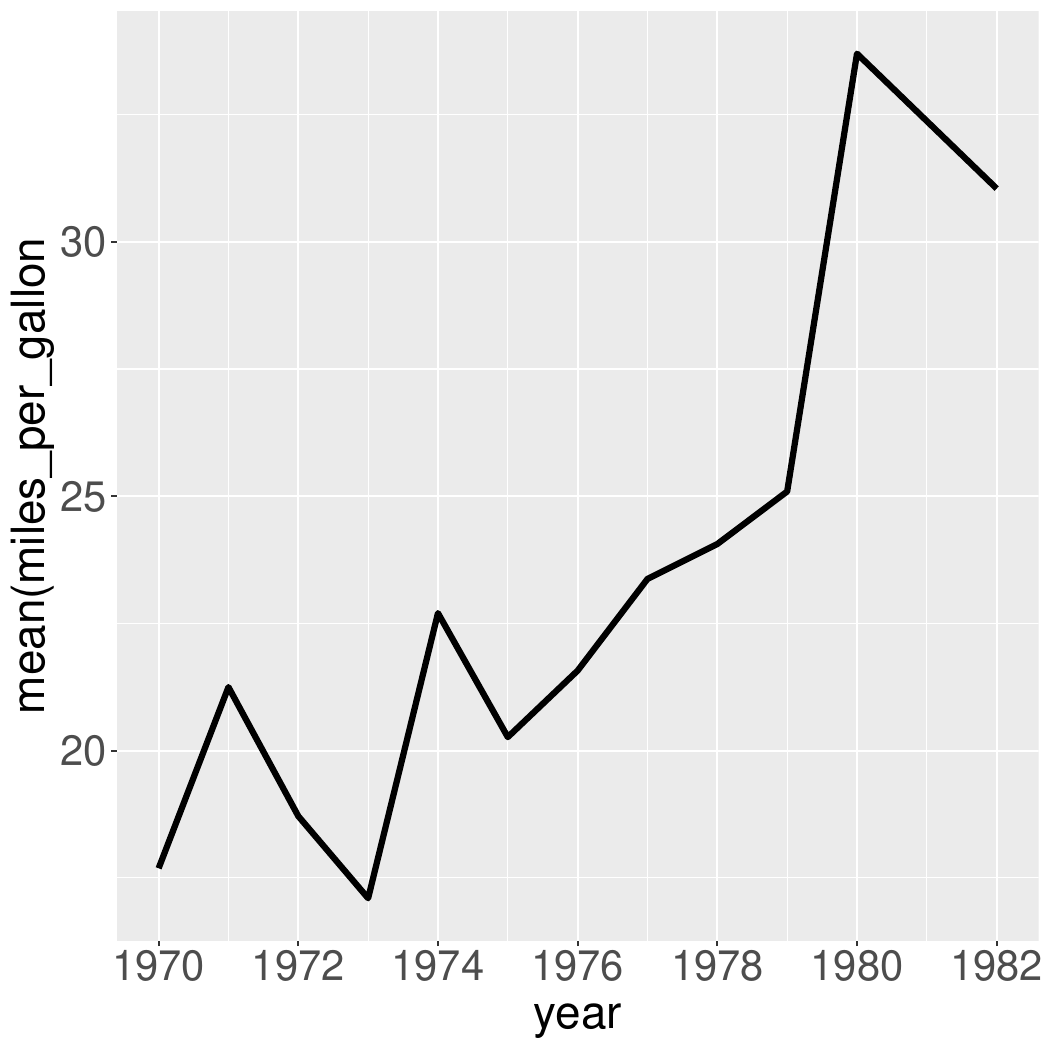}
\end{minipage}
\caption{SGL statement where a collective line geom represents multiple records in the post-CTA dataset.}
\label{fig:collective-geom}
\end{figure}

Within SGL's grammar, a geom is classified as individual if it represents each record in the post-CTA dataset by a distinct geometric object. Alternatively, a geom is classified
as collective if it represents multiple records in the post-CTA dataset by one geometric object.
In other words, individual geoms maintain a one-to-one relationship between post-CTA records and geometric objects, whereas collective geoms maintain a many-to-one relationship.
For example, points and lines are individual and collective geoms, respectively, as demonstrated in Figures \ref{fig:individual-geom} and \ref{fig:collective-geom} where the same data is represented using each.

\begin{figure}[H]
\centering
\begin{minipage}[t]{0.35\textwidth}
\vspace*{40pt}
\lstset{language=SQL, basicstyle=\ttfamily, columns=flexible}
\begin{lstlisting}
visualize
  age as x,
  circumference as y
from trees
using line;
\end{lstlisting}
\end{minipage}
\hfill
\begin{minipage}[t]{0.55\textwidth}
\vspace*{0pt}
\includegraphics[width=0.9\textwidth]{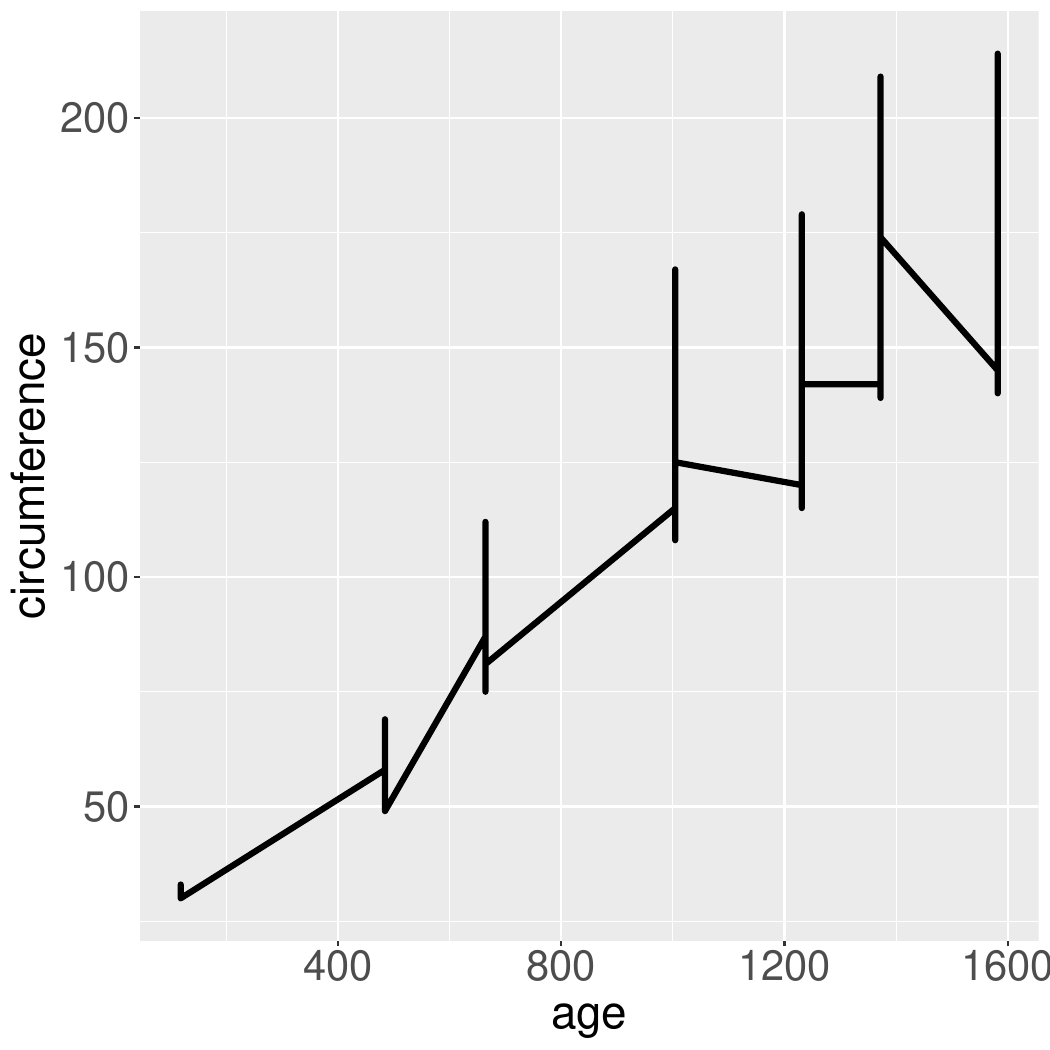}
\end{minipage}
\caption{SGL statement without a \texttt{collect by} clause, resulting in default collection behavior for the line geom.}
\label{fig:default-collection}
\end{figure}

\begin{figure}[H]
\centering
\begin{minipage}[t]{0.35\textwidth}
\vspace*{30pt}
\lstset{language=SQL, basicstyle=\ttfamily, columns=flexible}
\begin{lstlisting}
visualize
  age as x,
  circumference as y
from trees
collect by
  tree_id
using lines;
\end{lstlisting}
\end{minipage}
\hfill
\begin{minipage}[t]{0.55\textwidth}
\vspace*{0pt}
\includegraphics[width=0.9\textwidth]{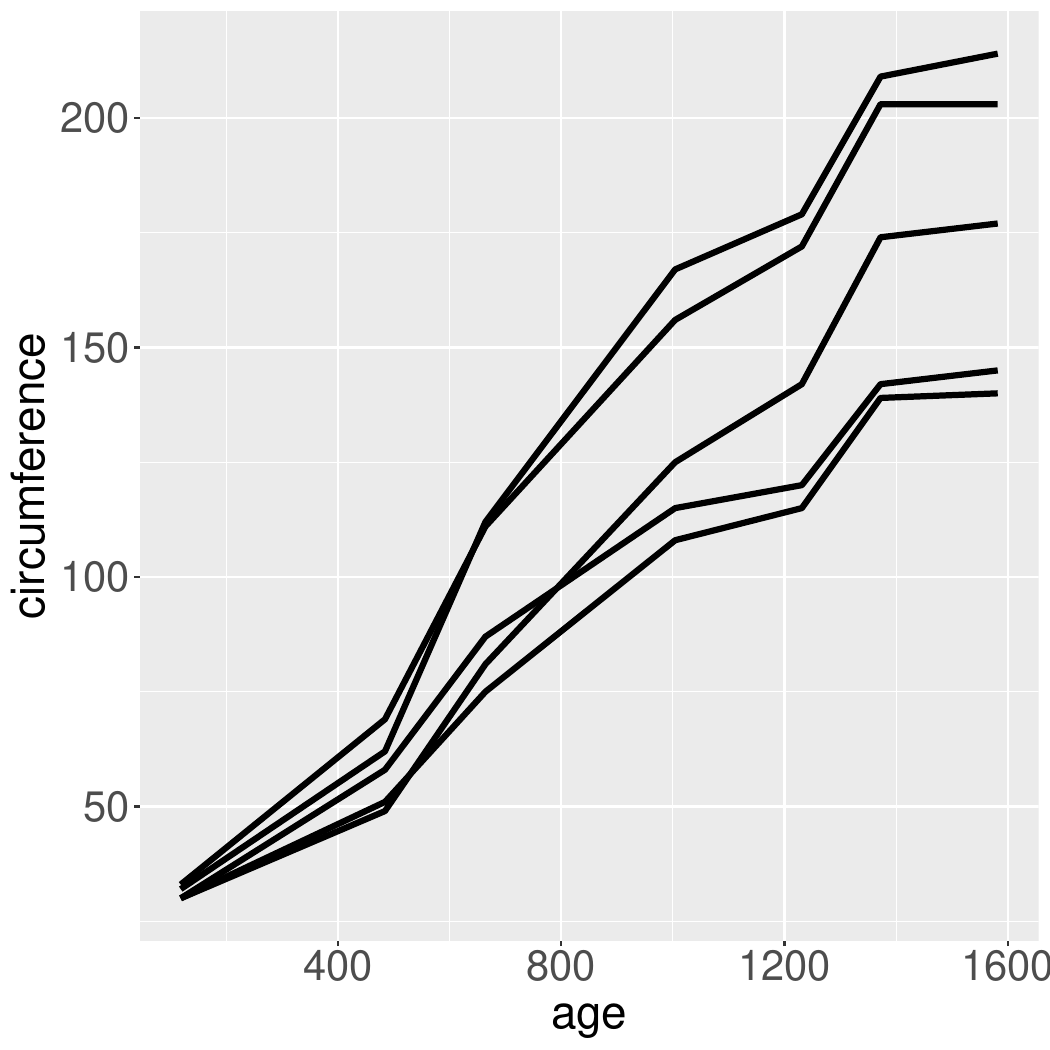}
\end{minipage}
\caption{SGL statement where default collection is overridden so that each tree is represented by a separate line.}
\label{fig:collect-by-clause}
\end{figure}

For collective geoms, the collection of records to represent by each object is determined implicitly according to reasonable defaults, as is common in grammar of graphics implementations \cite{wickham:2010, wilkinson:2005, vegalite:2017}.
This behavior can be overidden, however, by providing explicit collections in the \texttt{collect by} clause.
The \texttt{collect by} clause is similar to the \texttt{group by} clause, except that rather than defining groups to aggregate by, 
the \texttt{collect by} clause defines collections of records to be represented by one object.
Figures \ref{fig:default-collection} and \ref{fig:collect-by-clause} show default collection and overriding default collection via the \texttt{collect by} clause, respectively.

\subsection{Geom Qualifiers}
\begin{figure}[H]
\centering
\begin{minipage}[t]{0.35\textwidth}
\vspace*{38pt}
\lstset{language=SQL, basicstyle=\ttfamily, columns=flexible}
\begin{lstlisting}
visualize
  age as x,
  circumference as y
from trees
using regression line;
\end{lstlisting}
\end{minipage}
\hfill
\begin{minipage}[t]{0.55\textwidth}
\vspace*{0pt}
\includegraphics[width=0.9\textwidth]{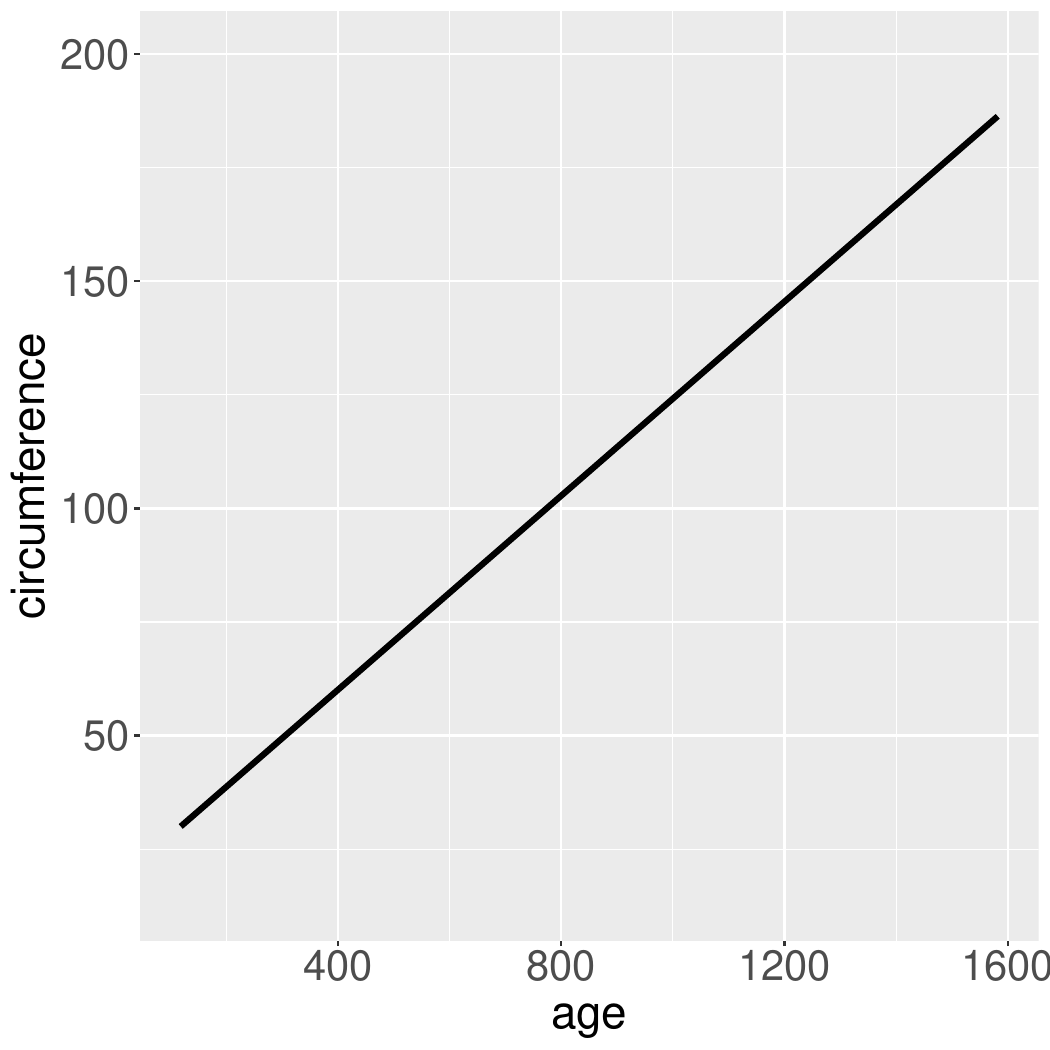}
\end{minipage}
\caption{SGL statement with a \texttt{regression} geom qualifier, resulting in a regression line of tree circumference against age.}
\label{fig:regression-qualifier}
\end{figure}

Geom qualifiers modify how geoms positionally represent data, and are specified via qualifier keywords that precede geom names within the \texttt{using} clause.
Geom qualifiers can largely be classified into two groups, statistical qualifiers and collision qualifiers. Statistical qualifiers modify the positional representation via statistical transformation, such
as linear regression, as shown in Figure \ref{fig:regression-qualifier}.

\begin{figure}[H]
\centering
\begin{minipage}[t]{0.35\textwidth}
\vspace*{40pt}
\lstset{language=SQL, basicstyle=\ttfamily, columns=flexible}
\begin{lstlisting}
visualize
  origin as x,
  miles_per_gallon as y
from cars
using points;
\end{lstlisting}
\end{minipage}
\hfill
\begin{minipage}[t]{0.55\textwidth}
\vspace*{0pt}
\includegraphics[width=0.9\textwidth]{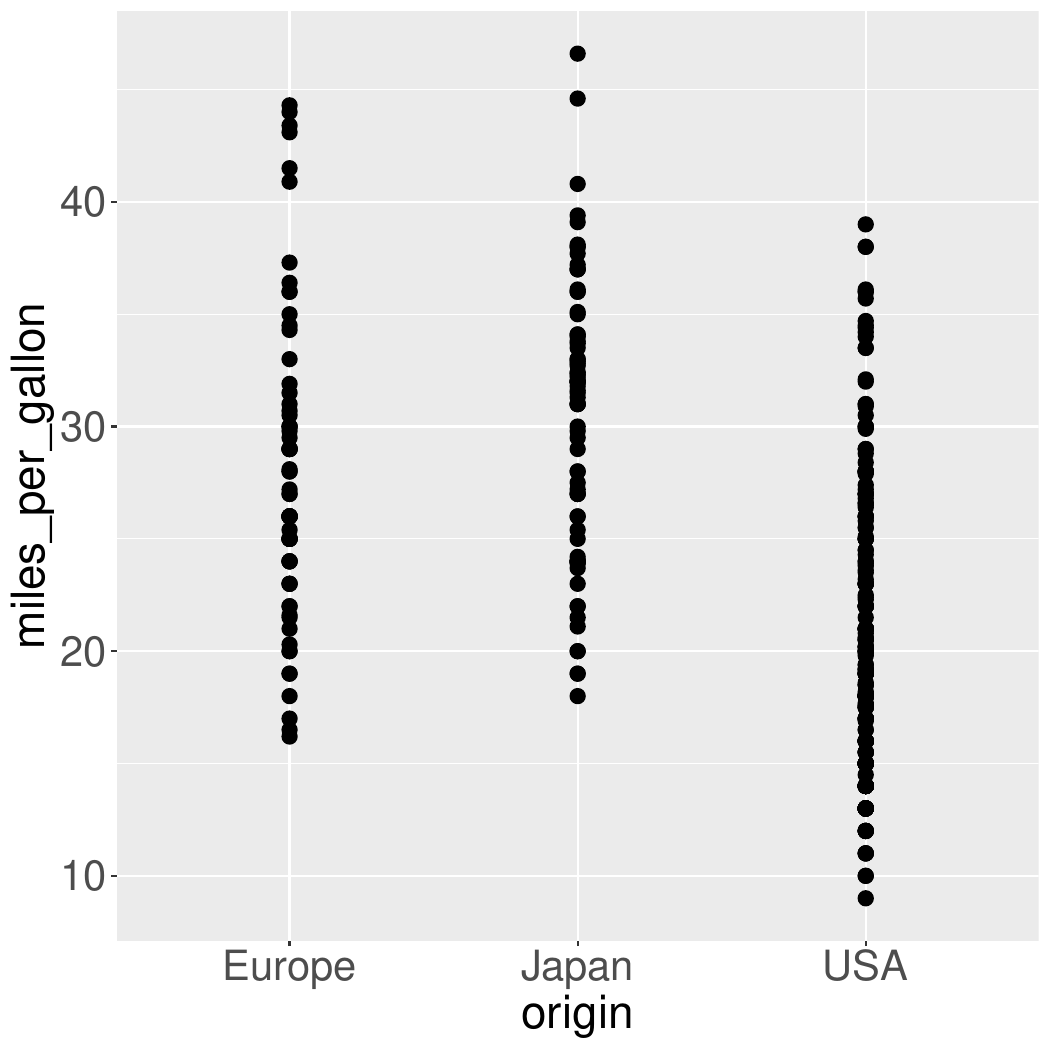}
\end{minipage}
\caption{SGL statement that generates a graphic with overlapping points that are not discernible.}
\label{fig:overlapping-points}
\end{figure}

\begin{figure}[H]
\centering
\begin{minipage}[t]{0.35\textwidth}
\vspace*{40pt}
\lstset{language=SQL, basicstyle=\ttfamily, columns=flexible}
\begin{lstlisting}
visualize
  origin as x,
  miles_per_gallon as y
from cars
using jittered points;
\end{lstlisting}
\end{minipage}
\hfill
\begin{minipage}[t]{0.55\textwidth}
\vspace*{0pt}
\includegraphics[width=0.9\textwidth]{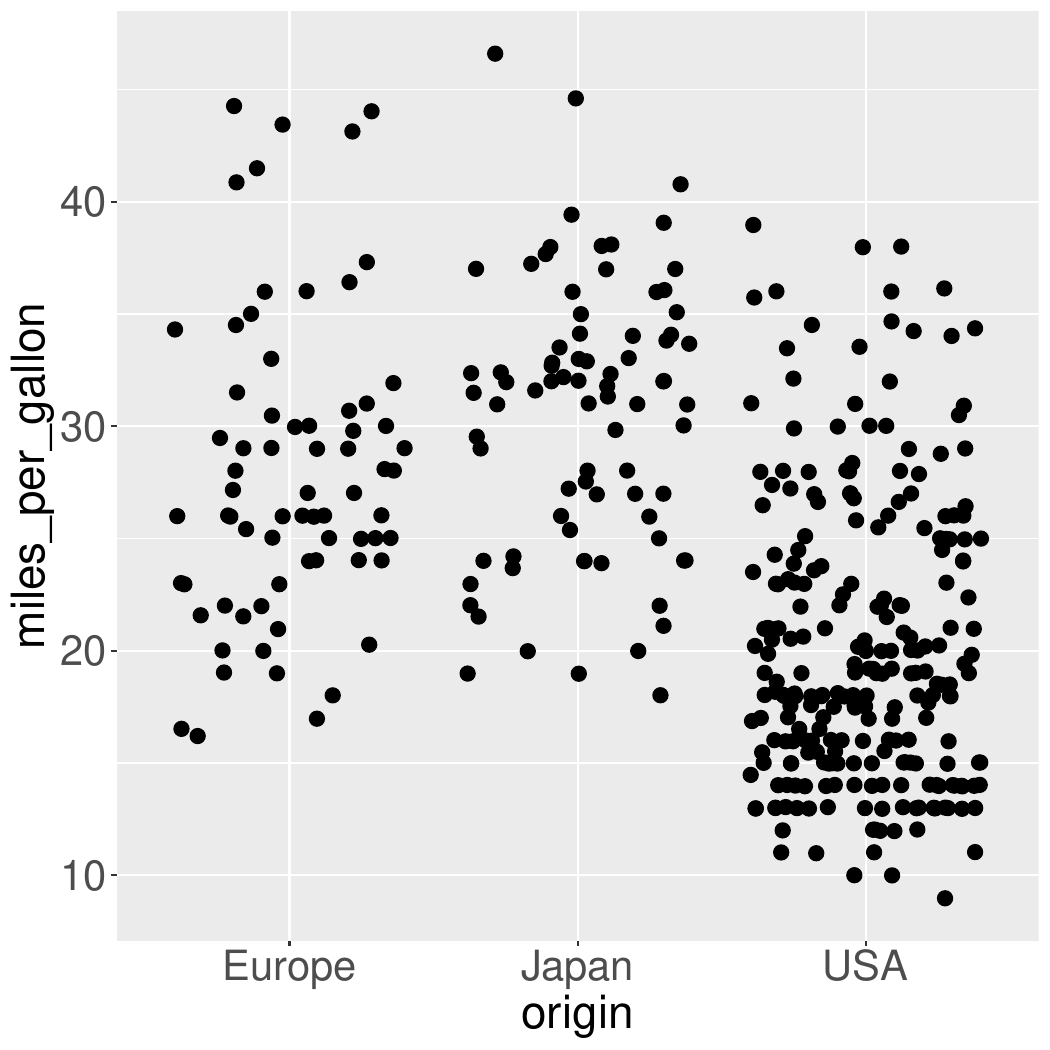}
\end{minipage}
\caption{SGL statement with a \texttt{jittered} geom qualifier that adds random variation so that overlapping points are discernible.}
\label{fig:jittered-qualifier}
\end{figure}

Collision qualifiers specify positional adjustments that are relevant to overlapping objects.
Figures \ref{fig:overlapping-points} and \ref{fig:jittered-qualifier} demonstrate using the \texttt{jittered} qualifier to add a small amount of random variation so that overlapping points are discernible.

\subsection{The Layer Operator}
\begin{figure}[H]
\centering
\begin{minipage}[t]{0.35\textwidth}
\vspace*{0pt}
\lstset{language=SQL, basicstyle=\ttfamily, columns=flexible}
\begin{lstlisting}
visualize
  horsepower as x,
  miles_per_gallon as y
from cars
using points

layer

visualize
  horsepower as x,
  miles_per_gallon as y
from cars
using regression line;
\end{lstlisting}
\end{minipage}
\hfill
\begin{minipage}[t]{0.55\textwidth}
\vspace*{5pt}
\includegraphics[width=0.9\textwidth]{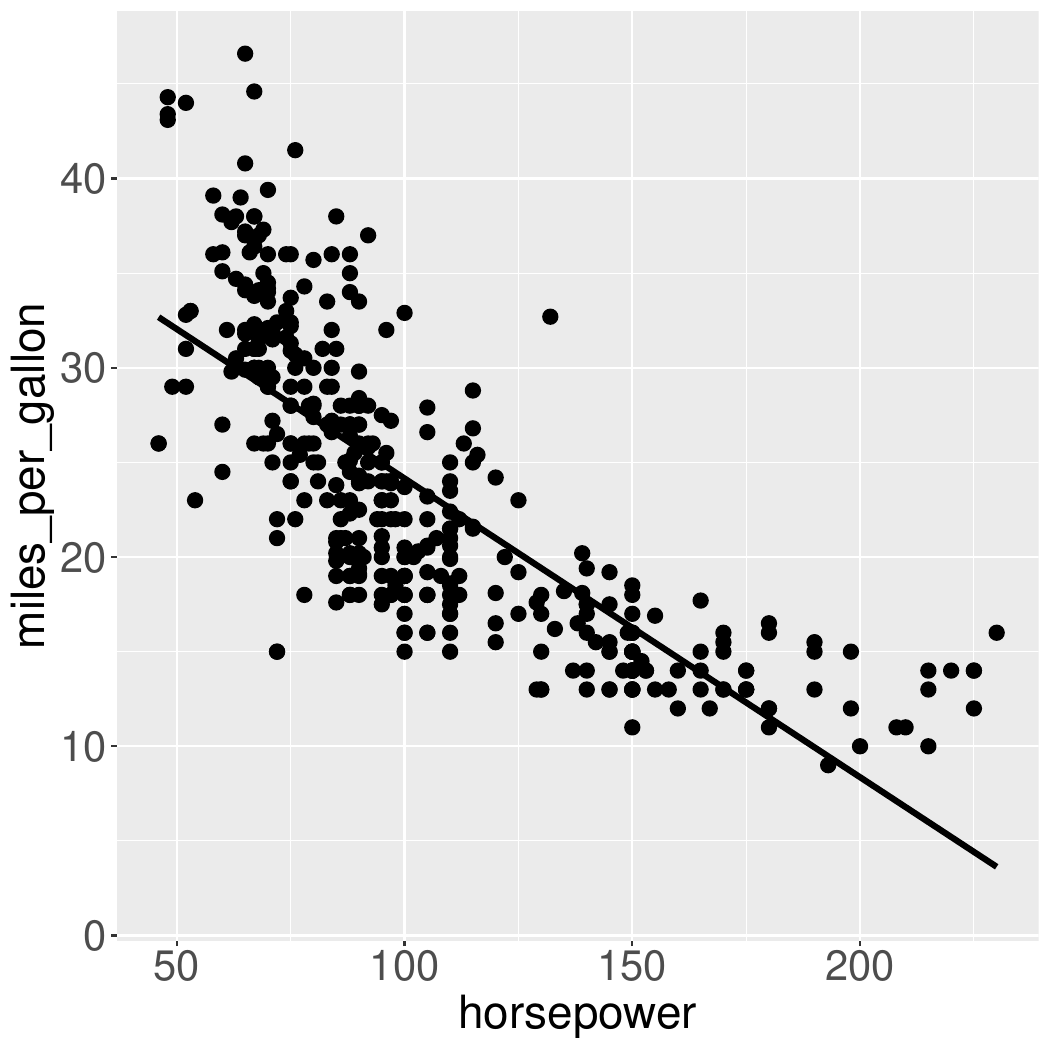}
\end{minipage}
\caption{SGL statement that uses the \texttt{layer} operator to add a regression line to a scatterplot.}
\label{fig:layer-operator}
\end{figure}

The graphics in previous sections contain a single layer of geometric objects.
It is common, however, to use multiple layers of objects to represent data in a single graphic.
In SGL, multiple layers can be defined and subsequently combined using the \texttt{layer} operator.
Figure \ref{fig:layer-operator} demonstrates layering a regression line on a scatterplot.

\begin{figure}[H]
\centering
\begin{minipage}[t]{0.35\textwidth}
\vspace*{26pt}
\lstset{language=SQL, basicstyle=\ttfamily, columns=flexible}
\begin{lstlisting}
visualize
  horsepower as x,
  miles_per_gallon as y
from cars
using (
  points
  layer
  regression line
);
\end{lstlisting}
\end{minipage}
\hfill
\begin{minipage}[t]{0.55\textwidth}
\vspace*{0pt}
\includegraphics[width=0.9\textwidth]{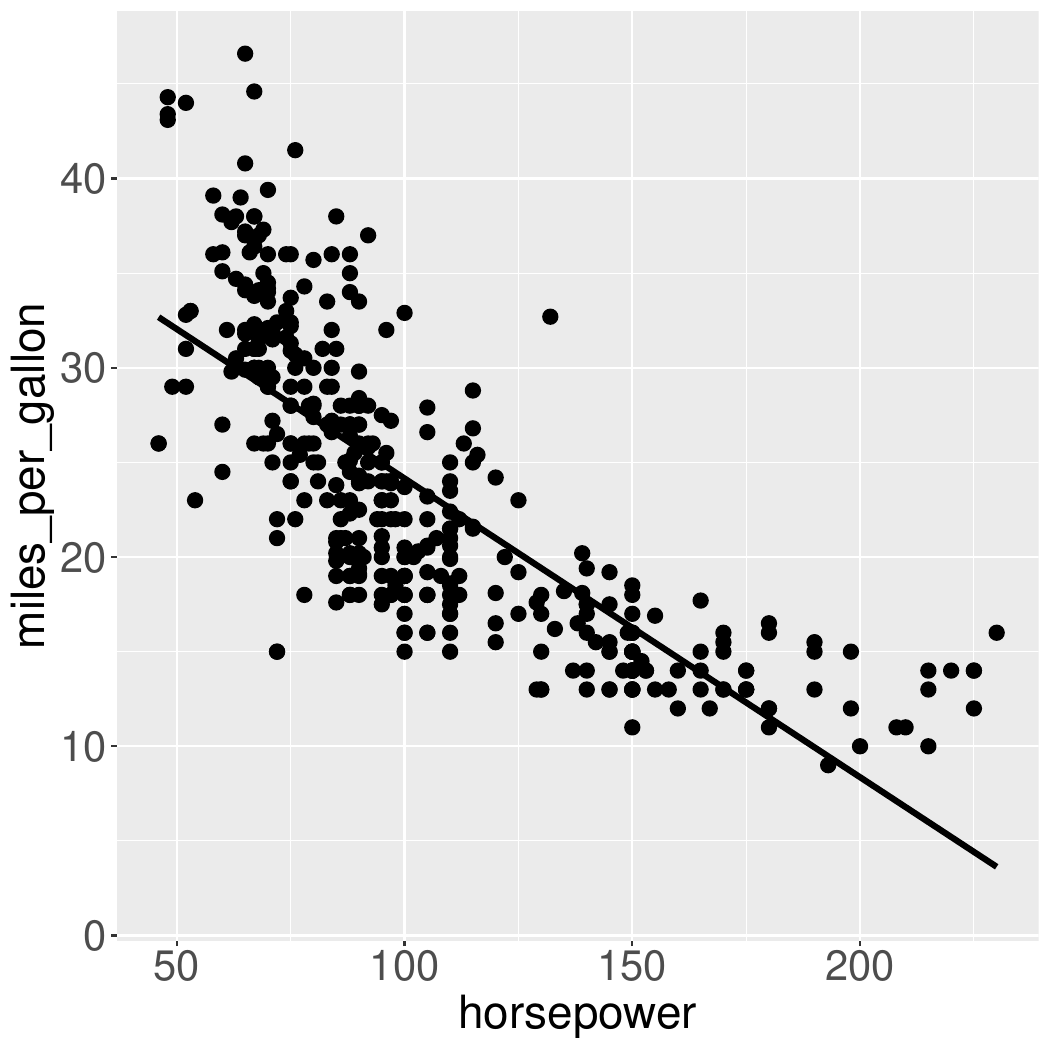}
\end{minipage}
\caption{SGL statement that uses layered geom expressions to reduce verbosity.}
\label{fig:layered-geom-expressions}
\end{figure}

Layers often share a data source and aesthetic mapping.
To reduce verbosity in such cases, the \texttt{layer} operator can be applied to geom expressions,
as shown in Figure \ref{fig:layered-geom-expressions}. A geom expression is either a geom name or a geom qualifier followed by a geom name.
 
Layers may have different data sources and aesthetic mappings. In SGL's grammar of graphics, however, a graphic has a single scale for each aesthetic, regardless of the number of layers.
As a result, a given aesthetic must be mapped to consistent type across all layers where it is present, e.g. an aesthetic cannot be mapped to a numerical type
in one layer and a categorical type in another.

\subsection{The Scale By Clause}
\label{sec:the-scale-by-clause}
\begin{figure}[H]
\centering
\begin{minipage}[t]{0.35\textwidth}
\vspace*{2pt}
\lstset{language=SQL, basicstyle=\ttfamily, columns=flexible}
\begin{lstlisting}
visualize
  horsepower as x,
  miles_per_gallon as y
from cars
using (
  points
  layer
  regression line
)
scale by
  log(x), log(y);
\end{lstlisting}
\end{minipage}
\hfill
\begin{minipage}[t]{0.55\textwidth}
\vspace*{0pt}
\includegraphics[width=0.9\textwidth]{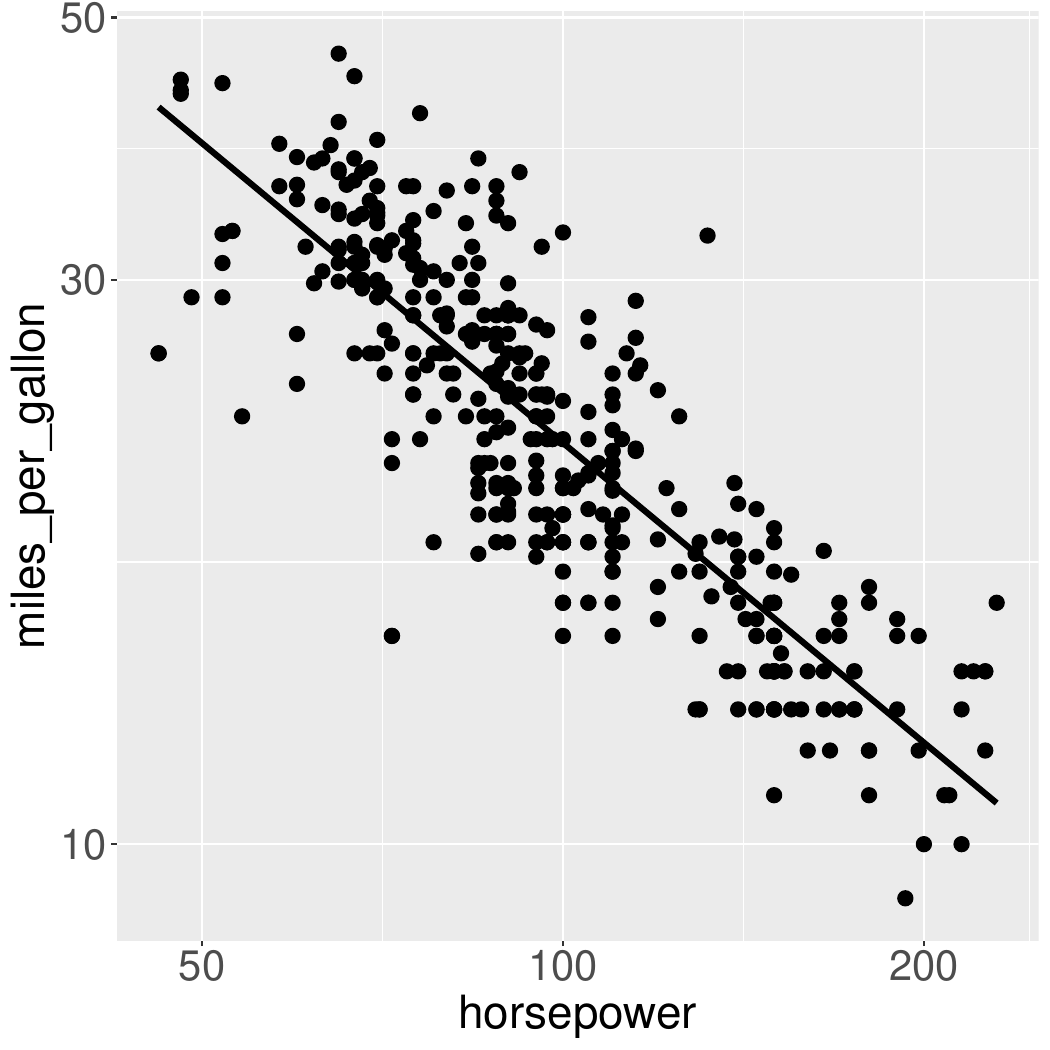}
\end{minipage}
\caption{SGL statement that specifies a \texttt{log} scale (base 10 by default) for the \texttt{x} and \texttt{y} aesthetics.}
\label{fig:log-scale}
\end{figure}

Each mapped aesthetic has a scale that determines how data values are mapped to the corresponding visual property.
Scales are determined implicitly by default, but can be explicitly specified within the \texttt{scale by} clause.
Figure \ref{fig:log-scale} demonstrates specifying a \texttt{log} scale for the \texttt{x} and \texttt{y} aesthetics, overriding the default linear scaling for numerical mappings.

As stated in section \ref{sec:column-level-transformations-and-aggregations}, scaling is performed prior to column-level transformations and aggregations.
Additionally, scaling is performed prior to any positional modifications specified by geom qualifiers, e.g. the regression calculation in Figure \ref{fig:log-scale} is
performed after log-scaling the x and y aesthetics.

\begin{figure}[H]
\centering
\begin{minipage}[t]{0.50\textwidth}
\vspace*{0pt}
\lstset{language=SQL, basicstyle=\ttfamily, columns=flexible}
\begin{lstlisting}
visualize
  log_hp as x,
  log_mpg as y
from (
  select
    log(horsepower) as log_hp,
    log(miles_per_gallon) as log_mpg
  from cars
)
using (
  points
  layer
  regression line
);
\end{lstlisting}
\end{minipage}
\hfill
\begin{minipage}[t]{0.40\textwidth}
\vspace*{30pt}
\includegraphics[width=0.9\textwidth]{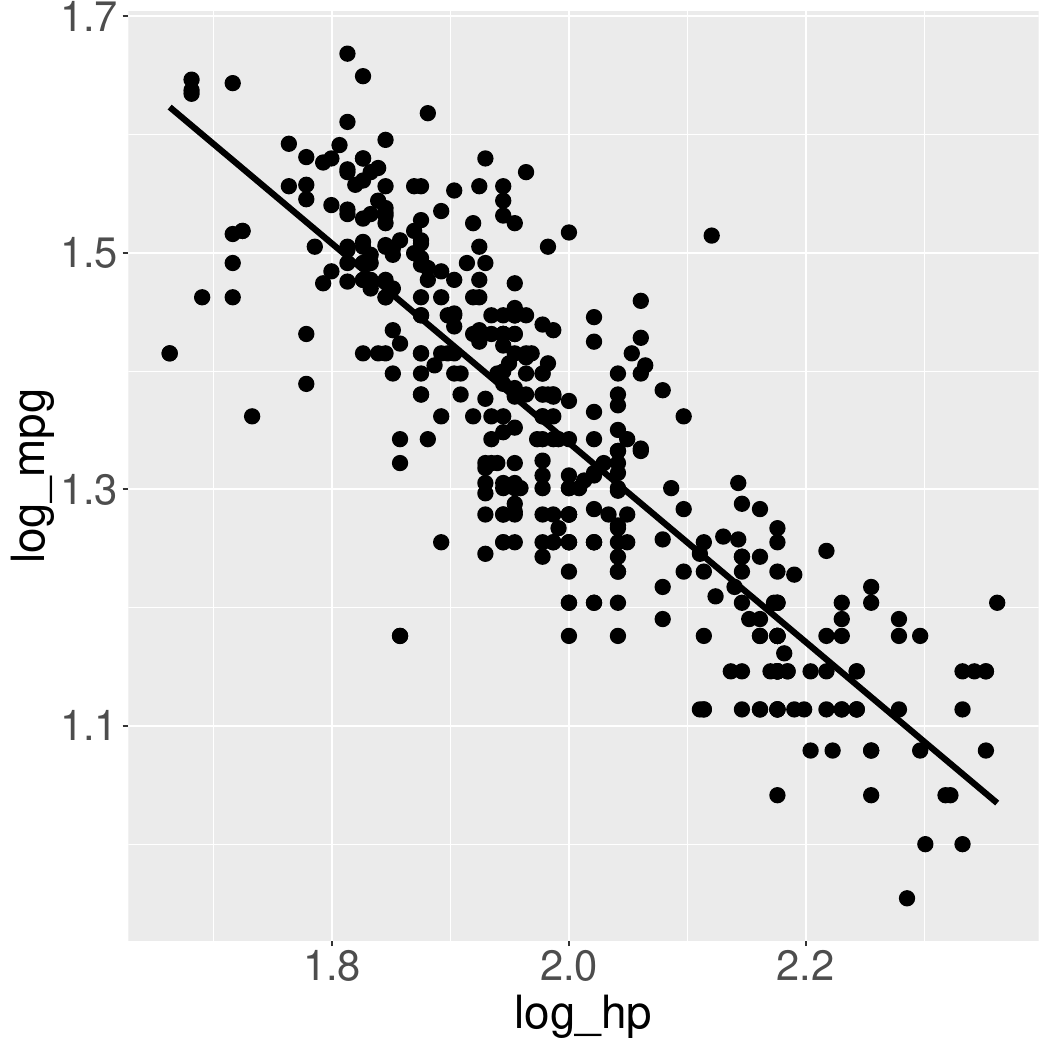}
\end{minipage}
\caption{SGL statement that applies a \texttt{log} function in a SQL subquery, resulting in modification of actual data values rather than modification of scales.}
\label{fig:log-values}
\end{figure}

Scaling functions such as \texttt{log} are applied to aesthetic names rather than column names as these are modifications to aesthetic scales rather than actual data values. 
This distinction is demonstrated in Figure \ref{fig:log-values}, where a \texttt{log} function is applied to actual data values in a SQL subquery.

\subsection{Coordinate Systems}
The graphics in prior sections use Cartesian coordinates, but alternative coordinate systems are valid.
In SGL, the coordinate system is inferred from the positional aesthetics referenced in the \texttt{visualize} clause(s).
For example, \texttt{x} and \texttt{y} aesthetics imply Cartesian coordinates, whereas
\texttt{theta} and \texttt{r} imply polar coordinates.

\begin{figure}[H]
\centering
\begin{minipage}[t]{0.35\textwidth}
\vspace*{32pt}
\lstset{language=SQL, basicstyle=\ttfamily, columns=flexible, showstringspaces=false}
\begin{lstlisting}
visualize
  count(*) as y,
  origin as color
from cars
group by
  origin
using bars;
\end{lstlisting}
\end{minipage}
\hfill
\begin{minipage}[t]{0.55\textwidth}
\vspace*{0pt}
\includegraphics[width=0.9\textwidth]{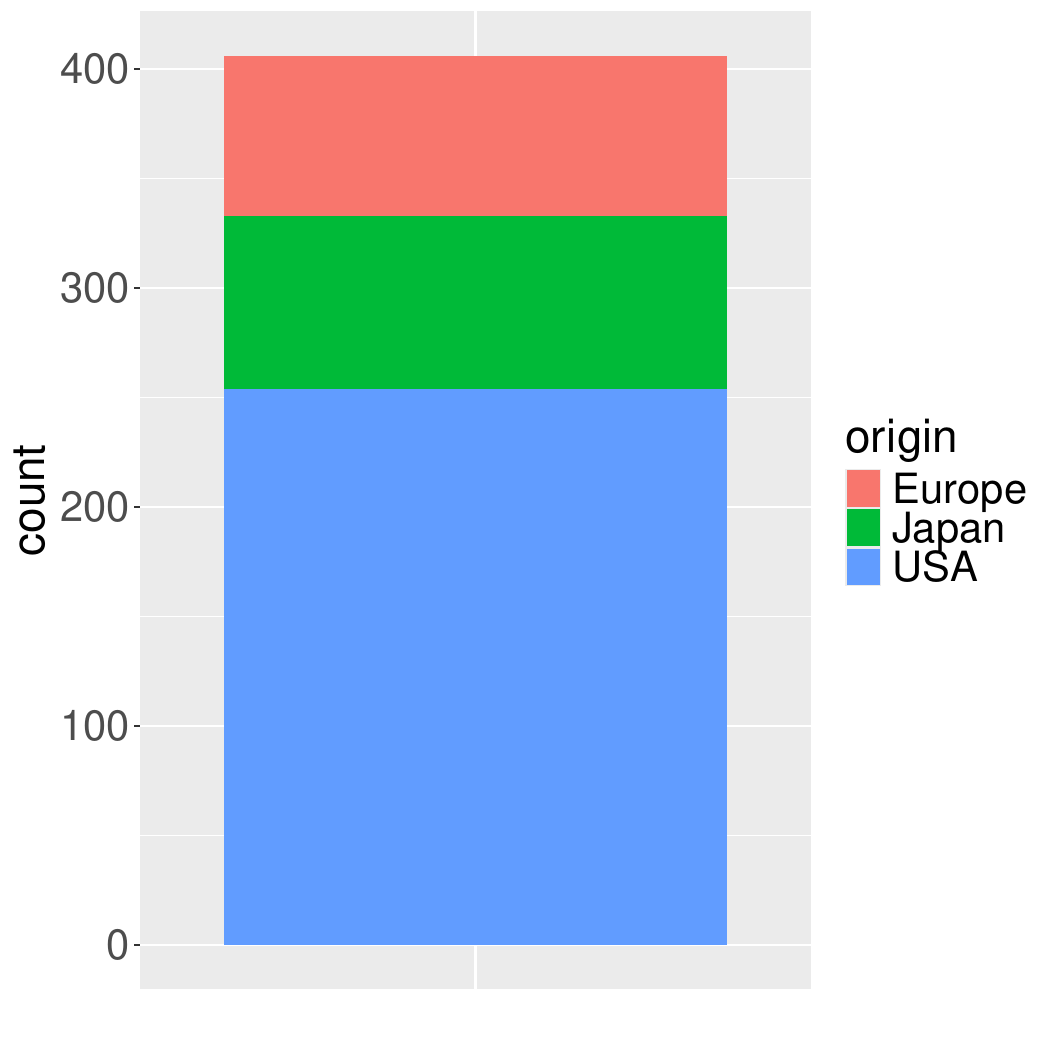}
\end{minipage}
\caption{SGL statement for a stacked bar chart in Cartesian coordinates.}
\label{fig:stacked-bar}
\end{figure}

\begin{figure}[H]
\centering
\begin{minipage}[t]{0.35\textwidth}
\vspace*{34pt}
\lstset{language=SQL, basicstyle=\ttfamily, columns=flexible, showstringspaces=false}
\begin{lstlisting}
visualize
  count(*) as theta,
  origin as color
from cars
group by
  origin
using bars;
\end{lstlisting}
\end{minipage}
\hfill
\begin{minipage}[t]{0.55\textwidth}
\vspace*{0pt}
\includegraphics[width=0.9\textwidth]{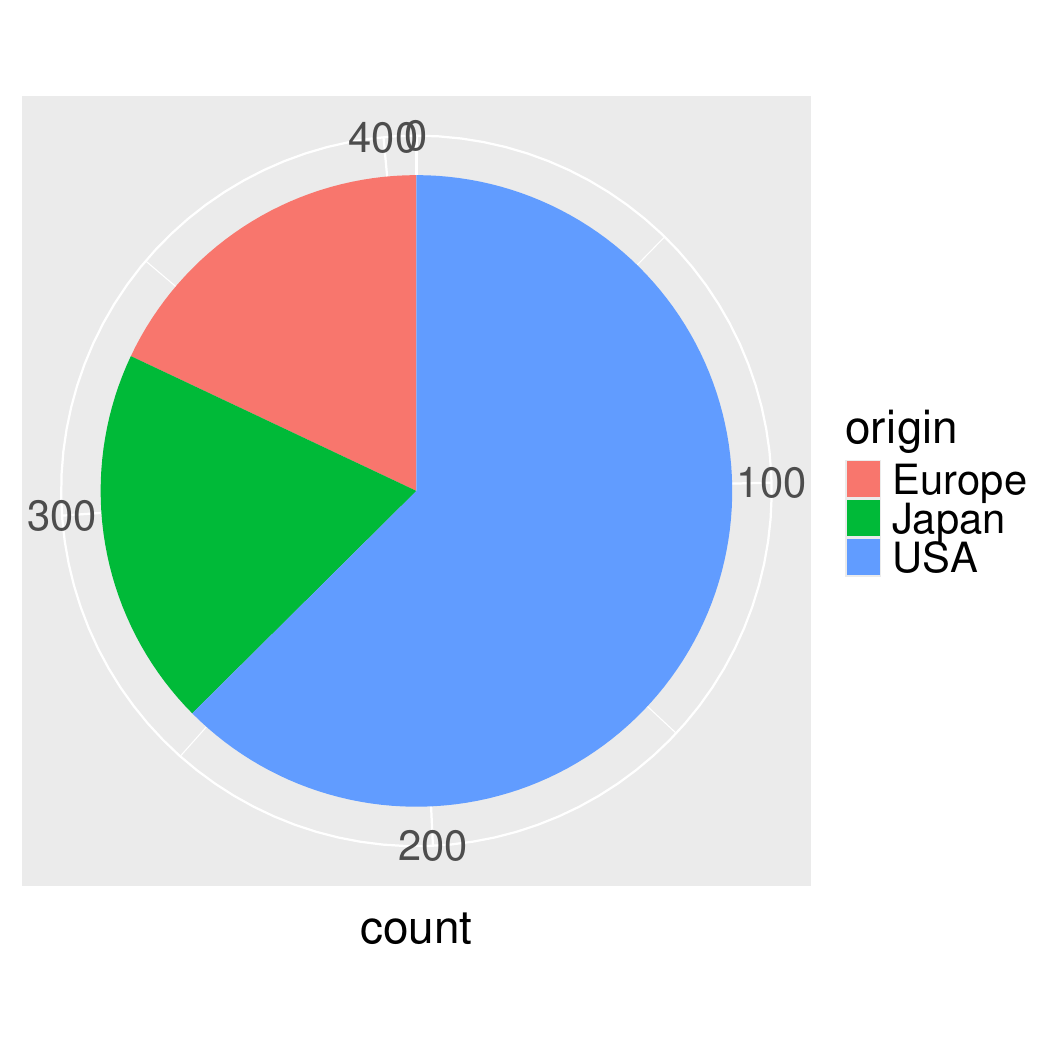}
\end{minipage}
\caption{SGL statement specifying a stacked bar chart in polar coordinates, resulting in what is commonly referred to as a pie chart.}
\label{fig:polar-coords}
\end{figure}

Figures \ref{fig:stacked-bar} and \ref{fig:polar-coords} display the same information in Cartesian and polar coordinates, respectively.
This example demonstrates the grammatical perspective that pie charts are stacked bar charts in a polar coordinate system \cite{wilkinson:2005, wickham:2010}.
In SGL, the bar geom is stacked by default, although this behavior can be overridden with the \texttt{unstacked} geom qualifier.

Since a SGL statement may have multiple layers, it may also have multiple \texttt{visualize} clauses, each with positional aesthetic mappings.
According to SGL's grammar of graphics, however, a graphic has a single coordinate system. As a consequence, the positional aesthetics referenced across layers must be consistent, e.g.
one layer cannot reference \texttt{x} and \texttt{y} aesthetics while another references \texttt{theta} and \texttt{r}.

\subsection{The Facet By Clause}

\begin{figure}[H]
\centering
\begin{minipage}[t]{0.35\textwidth}
\vspace*{34pt}
\lstset{language=SQL, basicstyle=\ttfamily, columns=flexible, showstringspaces=false}
\begin{lstlisting}
visualize
  horsepower as x,
  miles_per_gallon as y
from cars
using points
facet by
  origin;
\end{lstlisting}
\end{minipage}
\hfill
\begin{minipage}[t]{0.55\textwidth}
\vspace*{0pt}
\includegraphics[width=0.9\textwidth]{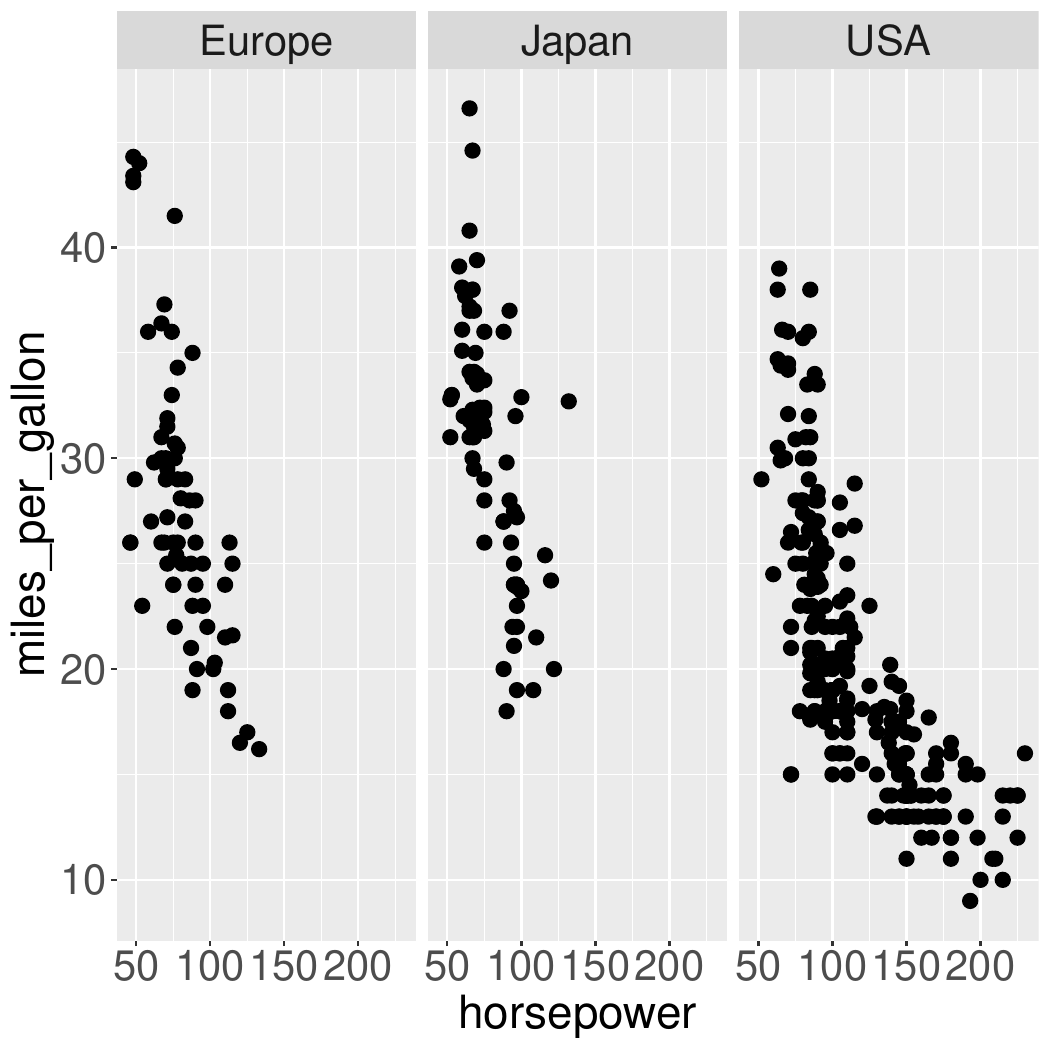}
\end{minipage}
\caption{SGL statement that facets by \texttt{origin}, resulting in a distinct panel for each country of origin.}
\label{fig:default-facet}
\end{figure}

\begin{figure}[H]
\centering
\begin{minipage}[t]{0.35\textwidth}
\vspace*{32pt}
\lstset{language=SQL, basicstyle=\ttfamily, columns=flexible, showstringspaces=false}
\begin{lstlisting}
visualize
  horsepower as x,
  miles_per_gallon as y
from cars
using points
facet by
  origin vertically;
\end{lstlisting}
\end{minipage}
\hfill
\begin{minipage}[t]{0.55\textwidth}
\vspace*{0pt}
\includegraphics[width=0.9\textwidth]{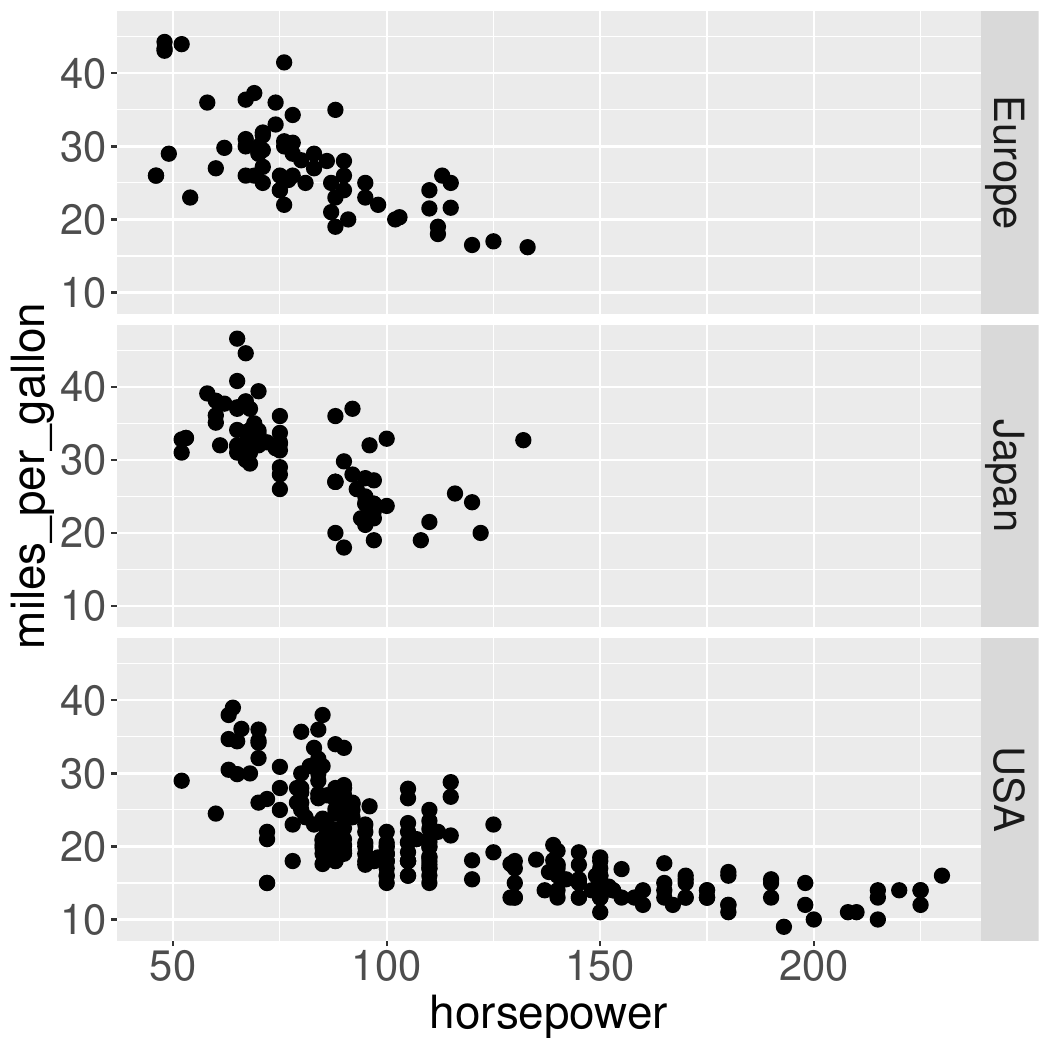}
\end{minipage}
\caption{SGL statement that specifies vertical panels using the \texttt{vertically} keyword.}
\label{fig:vertical-facet}
\end{figure}

Faceting generates small multiples where each panel represents a different partition of the source data.
Partitioning is determined by the unique values for expressions specified in the \texttt{facet by} clause.
Faceting by a single expression generates horizontal panels by default, but this can be modified
with an orientation keyword, as shown in Figures \ref{fig:default-facet} and \ref{fig:vertical-facet}.

\begin{figure}[H]
\centering
\begin{minipage}[t]{0.35\textwidth}
\vspace*{0pt}
\lstset{language=SQL, basicstyle=\ttfamily, columns=flexible, showstringspaces=false}
\begin{lstlisting}
visualize
  horsepower as x,
  miles_per_gallon as y
from (
  select
    *,
   case
     when year < 1977
     then '< 1977'
     else '>= 1977'
  end as 'era'
  from cars
)
using points
facet by
  era,
  origin;
\end{lstlisting}
\end{minipage}
\hfill
\begin{minipage}[t]{0.55\textwidth}
\vspace*{26pt}
\includegraphics[width=0.9\textwidth]{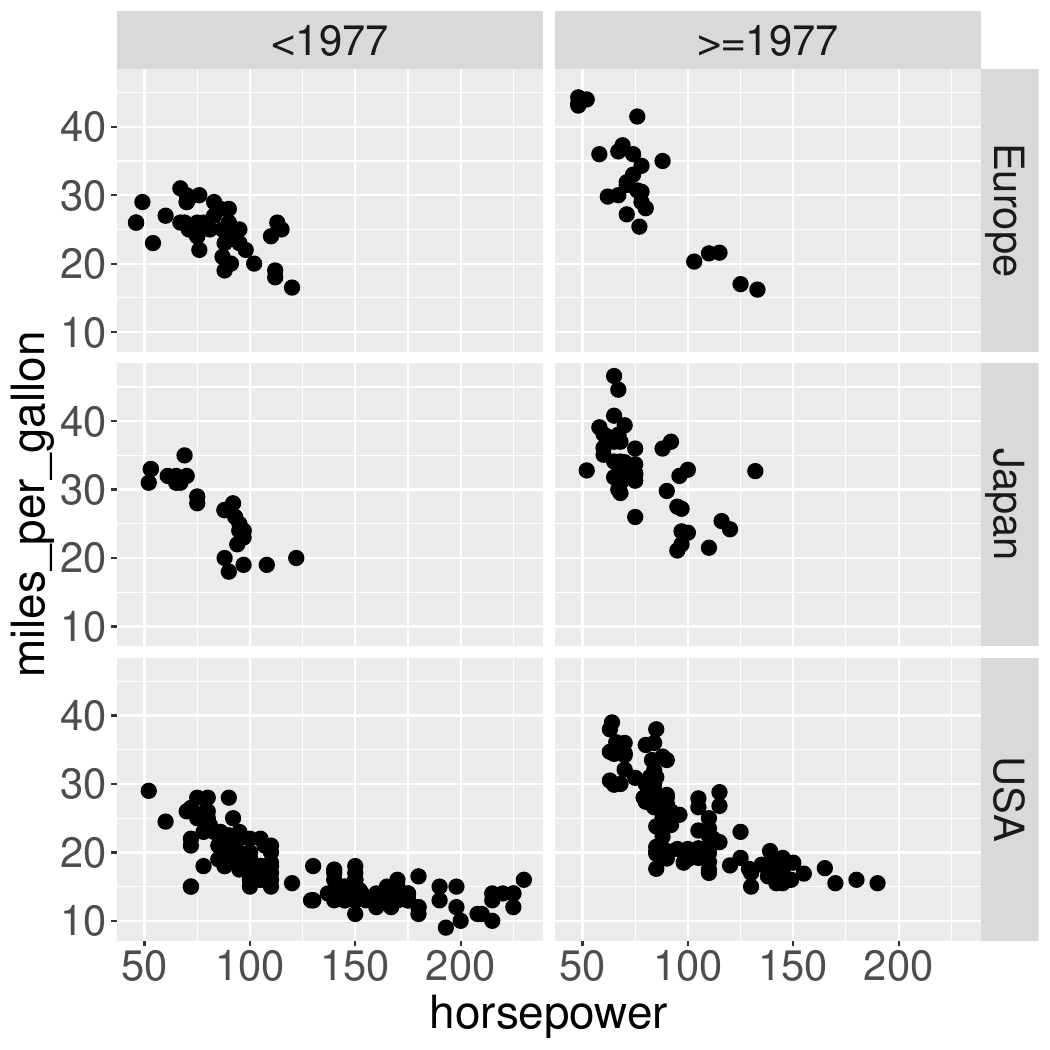}
\end{minipage}
\caption{SGL statement with two facet expressions, resulting in a grid of panels.}
\label{fig:multiple-facet-expressions}
\end{figure}

Two facet expressions may be specified, in which case one expression is represented horizontally while the 
other is represented vertically. This results in a grid representation of the Cartesian product of unique values,
as shown in Figure \ref{fig:multiple-facet-expressions}.

In SGL's grammar of graphics, facets are a graphic-level property, meaning that each SGL statement has at
most one \texttt{facet by} clause, and that each layer is faceted accordingly.

\subsection{The Title Clause}

\begin{figure}[H]
\centering
\begin{minipage}[t]{0.35\textwidth}
\vspace*{28pt}
\lstset{language=SQL, basicstyle=\ttfamily, columns=flexible, showstringspaces=false}
\begin{lstlisting}
visualize
  horsepower as x,
  miles_per_gallon as y
from cars
using points
title
  x as 'Horsepower',
  y as 'Miles Per Gallon';
\end{lstlisting}
\end{minipage}
\hfill
\begin{minipage}[t]{0.55\textwidth}
\vspace*{0pt}
\includegraphics[width=0.9\textwidth]{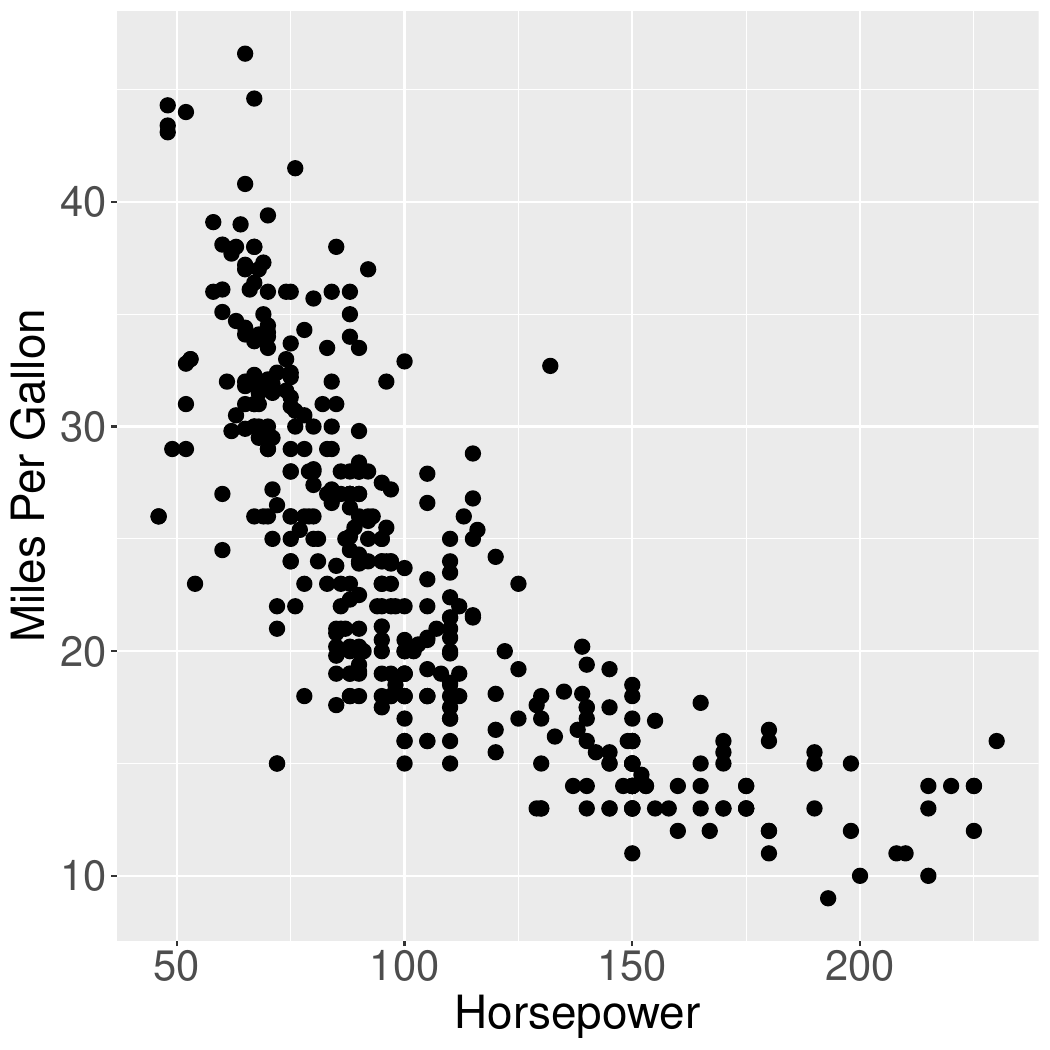}
\end{minipage}
\caption{SGL statement with explicit titles specified within the \texttt{title} clause.}
\label{fig:explicit-title}
\end{figure}

Titles for aesthetic scales are automatically determined from corresponding aesthetic mappings.
This can be overridden, however, by providing explicit titles in the \texttt{title} clause,
as shown in Figure \ref{fig:explicit-title}.

\section{SGL's Grammar of Graphics}
To describe SGL's grammar of graphics, we will assume familiarity with its closest predecessor, the layered grammar of graphics \cite{wickham:2010},
and elaborate on key distinctions between the two.

Figure \ref{fig:high-level-grammar-mapping} shows a mapping of graphical components where the details of the layer concept have been abstracted away.

\begin{figure}[H]
\centering
\begin{tikzpicture}[
    node distance=0.4cm,
    label/.style={font=\normalsize, anchor=west},
    sublabel/.style={font=\normalsize, anchor=west}
]

\node[label] (ggplot_header) at (0,0) {Layered Grammar};
\node[label] (ggplot_defaults) at (0,-0.6) {Defaults};
\node[label] (ggplot_layers) at (0,-1.2) {Layers};
\node[label] (ggplot_facets) at (0,-1.8) {Facets};
\node[label] (ggplot_scales) at (0,-2.4) {Scales};
\node[label] (ggplot_coordinates) at (0,-3.0) {Coordinate System};

\node[sublabel] (sgl_header) at (5,0) {SGL's Grammar};
\node[sublabel] (sgl_layers) at (5,-1.2) {Layers};
\node[sublabel] (sgl_facets) at (5,-1.8) {Facets};
\node[sublabel] (sgl_scales) at (5,-2.4) {Scales};
\node[sublabel] (sgl_coordinates) at (5,-3.0) {Coordinate System};

\draw (0, -0.25) - - (8.5, -0.25);
\draw[->, thick] (ggplot_layers.east) -- (sgl_layers.west);
\draw[->, thick] (ggplot_facets.east) -- (sgl_facets.west);
\draw[->, thick] (ggplot_scales.east) -- (sgl_scales.west);
\draw[->, thick] (ggplot_coordinates.east) -- (sgl_coordinates.west);

\end{tikzpicture}
\caption{Mapping of components between the layered grammar and SGL's grammar.}
\label{fig:high-level-grammar-mapping}
\end{figure}

\begin{figure}[H]
\centering
\begin{minipage}[t]{0.35\textwidth}
\vspace*{0pt}
\lstset{language=R, basicstyle=\ttfamily, columns=flexible, showstringspaces=false}
\begin{lstlisting}
ggplot(
  cars,
  aes(
    x=horsepower,
    y=miles_per_gallon
  )
) +
geom_point() +
geom_smooth(
  method='lm',
  se=FALSE,
  color='black'
)
\end{lstlisting}
\end{minipage}
\hfill
\begin{minipage}[t]{0.55\textwidth}
\vspace*{2pt}
\includegraphics[width=0.9\textwidth]{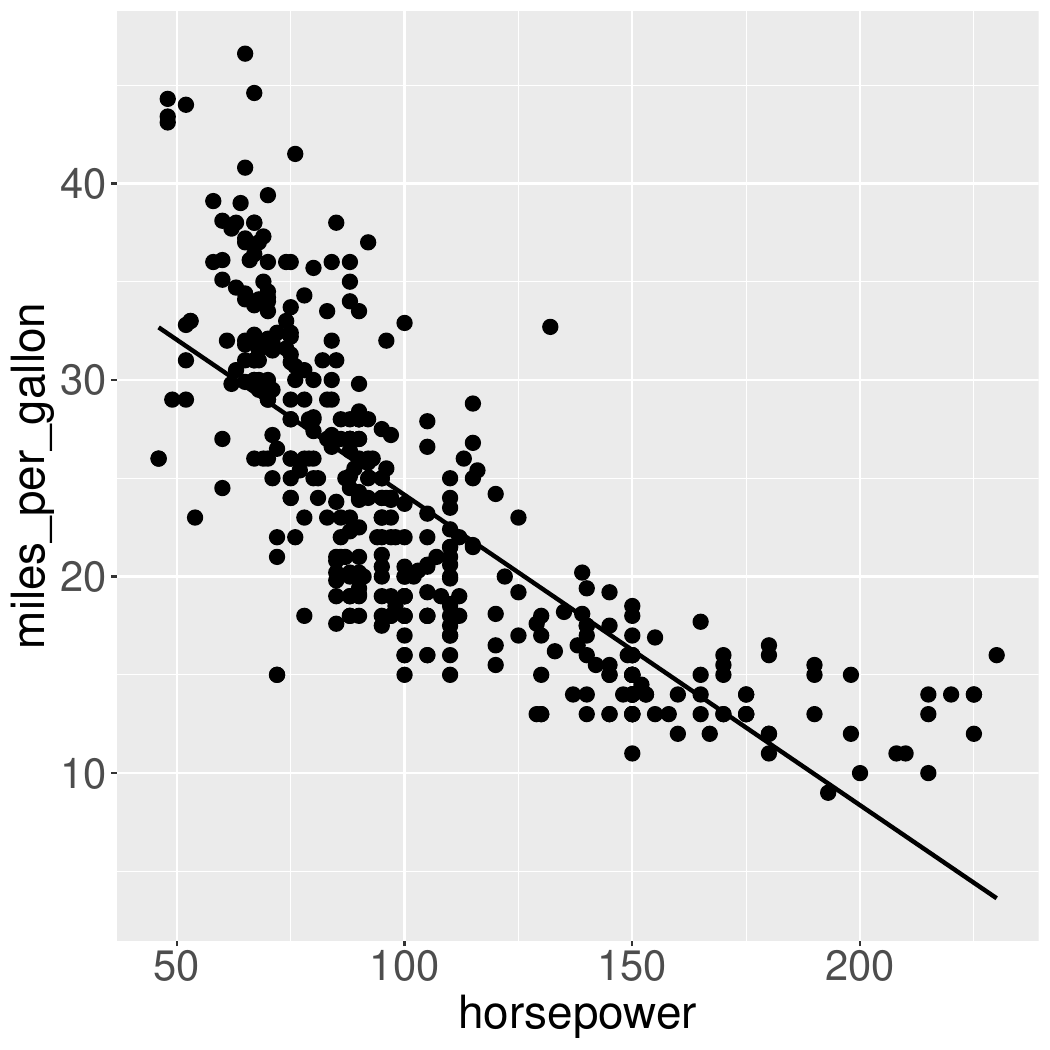}
\end{minipage}
\caption{ggplot2 specification where a default dataset and mapping is used across multiple layers.}
\label{fig:ggplot2-defaults}
\end{figure}

In the layered grammar, a default dataset and aesthetic mapping may be provided for a graphic, with layers having the option to override these defaults.
This leads to more concise specification since a dataset and mapping need not to be repeated,
as shown in the ggplot2 specification of Figure \ref{fig:ggplot2-defaults}.

In SGL's grammar, there is no concept of a default dataset or mapping for a graphic.
Instead, SGL enables concise specification of layers that share the same dataset and mapping through
layered geom expressions, as demonstrated in Figure \ref{fig:layered-geom-expressions}.
This syntax does not support a more concise specification for all default and override scenarios supported by the layered grammar,
but it better aligns with the desired SQL aesthetic of the language.

The two grammars are in accordance regarding facets, scales, and coordinate system.
There are differences, however, in the components of layers, as shown in the component mapping of Figure \ref{fig:layer-grammar-mapping}.

\begin{figure}[H]
\centering
\begin{tikzpicture}[
    node distance=0.4cm,
    label/.style={font=\normalsize, anchor=west},
    sublabel/.style={font=\normalsize, anchor=west}
]

\node[label] (ggplot_header) at (0,0) {Layered Grammar};
\node[label] (ggplot_data) at (0,-0.6) {Data};
\node[label] (ggplot_mapping) at (0,-1.2) {Mapping};
\node[label] (ggplot_grouping) at (0,-1.6) {*Group};
\node[label] (ggplot_geom) at (0,-3.0) {Geom};
\node[label] (ggplot_stat) at (0,-3.6) {Stat};
\node[label] (ggplot_position) at (0,-4.8) {Position};

\node[sublabel] (sgl_header) at (5,0) {SGL's Grammar};
\node[sublabel] (sgl_data) at (5,-0.6) {Data};
\node[sublabel] (sgl_mapping) at (5,-1.2) {Mapping};
\node[sublabel] (sgl_grouping) at (5,-1.8) {Groupings};
\node[sublabel] (sgl_collection) at (5,-2.4) {Collections};
\node[sublabel] (sgl_geom) at (5,-3.0) {Geom};
\node[sublabel] (sgl_column_level_transformation) at (5,-3.6) {Column-Level Transformations};
\node[sublabel] (sgl_aggregation) at (5,-4.2) {Aggregations};
\node[sublabel] (sgl_geom_qualifier) at (5,-4.8) {Geom Qualifiers};

\draw (0, -0.25) - - (8.5, -0.25);
\draw[->, thick] (ggplot_data.east) -- (sgl_data.west);
\draw[->, thick] (ggplot_mapping.east) -- (sgl_mapping.west);
\draw[->, thick] (ggplot_grouping.east) -- (sgl_grouping.west);
\draw[->, thick] (ggplot_grouping.east) -- (sgl_collection.west);
\draw[->, thick] (ggplot_geom.east) -- (sgl_geom.west);
\draw[->, thick] (ggplot_stat.east) -- (sgl_column_level_transformation.west);
\draw[->, thick] (ggplot_stat.east) -- (sgl_aggregation.west);
\draw[->, thick] (ggplot_stat.east) -- (sgl_geom_qualifier.west);
\draw[->, thick] (ggplot_position.east) -- (sgl_geom_qualifier.west);

\end{tikzpicture}
\caption{Mapping of layer components between the layered grammar and SGL's grammar.}
\footnotesize{$^*$Group is defined via an aesthetic mapping in the layered grammar.}
\label{fig:layer-grammar-mapping}
\end{figure}
In the layered grammar, each layer has one associated statistical transformation, or stat.
Stats cover a wide array of transformations. In SGL's grammar, these stats are broken down into
column-level transformations, aggregations, and  geom qualifiers.
SGL's grammar allows for the combination of these components in a single layer,
i.e., column-level transformations, aggregations, and statistical transformation via geom qualification
can be specified together within a single layer. In such cases, column-level transformations are performed first,
followed by aggregations, followed by statistical transformation.
This provides for more flexibility than the layered grammar, which requires a single
stat for each layer.
However, the primary motivation for refining the stat concept in SGL's grammar was not the additional flexibility,
but rather that such a grammar maps naturally to language components with analogs in SQL, such as
transformation functions applied to columns, aggregation functions, and a \texttt{group by} clause.
In other words, refining the stat concept in this way contributes to the SQL aesthetic of the SGL language.

In the layered grammar, a mapping for a group aesthetic may be provided,
in which case the groupings for aggregation statistics and collections for collective geoms
are both derived from that single group mapping.
In SGL's grammar, aggregation groupings and collections are distinct components
that may be modified independently, allowing for additional flexibility.

Lastly, positional adjustments (Position in Figure \ref{fig:layer-grammar-mapping}) and stats that aren't column-level transformations or aggregations
combine to form geom qualifiers in SGL's grammar.
These are combined due to the shared characteristic of modifying how geoms positionally represent data, and because the combination lends itself
to a uniform syntactic treatment that contributes to the SQL aesthetic.

The two grammars are in accordance regarding the data, geom, and mapping components,
aside from the group aesthetic of the mapping component previously discussed.

\section{Conclusion}
This paper presented SGL, a graphics language that is aesthetically similar to SQL.
Additionally, it presented SGL's underlying grammar of graphics, which has been adapted from prior grammars to support the SQL-like features of the language.
The SQL aesthetic as well as the concise yet expressive nature derived from the grammatical foundation
make SGL a suitable graphical counterpart to SQL, enabling specification of statistical graphics within SQL query interfaces.

\bibliographystyle{plain}
\bibliography{refs}

\end{document}